\documentclass[12pt]{article}
\usepackage{epsfig}
\usepackage{epstopdf}
\usepackage{amsmath}
\usepackage{latexsym}
\usepackage{textcomp}

\usepackage{amssymb}
\usepackage{amsfonts}
\usepackage{amsmath}
\usepackage[all]{xy}
\usepackage{amsfonts}
\usepackage{amssymb,amsmath}
\usepackage{hyperref}
\usepackage{natbib}

\usepackage[margin=1in]{geometry}
\usepackage[singlespacing]{setspace}

\newtheorem{definition}{ {Definition}}[section]
\newtheorem{theorem}[definition]{\bf {THEOREM}}
\newtheorem{lemma}[definition]{\bf {Lemma}}

\newcommand{\proof}{\noindent{\bf Proof: }}

 \bibpunct[, ]{(}{)}{,}{a}{}{,}%
\usepackage{graphicx,hyperref}
\usepackage{epsfig}
\usepackage{epstopdf}
\newcommand{\rar}{\rightarrow}
\newcommand{\Xomit}[1]{}

\begin{document}

\title{Searching and Bargaining with Middlemen}

\author{Th\`anh Nguyen\footnote{t-nguyen@kellogg.northwestern.edu},  \hspace{3mm} Vijay G. Subramanian\footnote{v-subramanian at northwestern.edu}, \hspace{3mm} Randall A. Berry\footnote{rberry@ece.northwestern.edu }\\
Northwestern University, Evanston IL 60208}

\maketitle
\onehalfspacing

\begin{abstract}
We study  decentralized markets with the presence of middlemen, modeled by  a non-cooperative bargaining game  in  trading networks. Our goal is to investigate  how the network structure of the market and the role of middlemen  influence the market's  efficiency and  fairness. We introduce the concept of limit stationary equilibrium in a general trading network and use it to analyze how competition among middlemen is influenced by the network structure, how endogenous  delay emerges in trade and  how surplus is shared between producers and consumers.
\end{abstract}

\section{Introduction}\label{intro}
In most markets trade does not involve just producers and consumers but also one or more middlemen serving as intermediaries.  For example, brokers and market makers fill this role in financial markets as do wholesalers and retailers in many manufacturing industries. Classical economic approaches to studying markets, such as competitive equilibrium analysis, largely abstract away the role of such middlemen; a point made in the introduction of \cite{RubinsteinWolinsky1987}, who attribute this is to a lack of modeling how trade occurs and the associated frictions involved. \cite{RubinsteinWolinsky1987} offer a solution to this shortcoming by adopting a search theoretic model as in \cite{diamond1979equilibrium,mortensen1982matching,diamond1982wage}.  Agents meet pairwise over time and must wait until they meet a suitable partner to trade. The time it takes to find a partner is costly thus introducing a search friction. The role of middlemen is in reducing this friction. Subsequently there has been much work in studying different models of trade (e.g., various non-cooperative bargaining models) and using these to analyze how middlemen influence the formation of prices and the efficiency of trade. 

Much of the aforementioned work has focused on models in which all producers and consumers have access to the same middlemen. However, often this is not the case due, for example, to various institutional or physical barriers. One example of this as pointed out by \cite{blume2009trading} is in agricultural supply chains of developing countries. In such cases, due to inadequate transportation infrastructure, farmers may only be able to trade in local markets. Such relationships are naturally modeled via a network. \cite{blume2009trading} consider such trading networks with a focus on characterizing how network structure effects equilibrium prices set by middlemen which have full information and full bargaining power and so there are no trading frictions.  Similar equilibrium questions have also been studied in the supply chain literature (e.g.~\cite{nagurney2002supply}). 

The first line of work described above focuses on modeling trade and its associated frictions, assuming simple trading networks (often with a single middleman). On the other hand, the second line of work focuses on the impact of network structure but does not account for trade frictions. 
The interaction of both these effects is not well understood. In a more complex network, the search problem facing an agent will depend on her location in the network, and the presence of such frictions will naturally given rise to different equilibria than in models such as \cite{blume2009trading}. 

This paper provides a starting point to bridge this gap: as in \cite{blume2009trading} we consider a trading network connecting consumers to producers, but as in \cite{RubinsteinWolinsky1987},  agents randomly meet over time and engage in  non-cooperative bargaining protocols.  Thus, our  paper provides a general framework that incorporates three important features of markets. First is the underlying network  structure: not all pairs of agents can interact in the market. The second is the non-cooperative bargaining setting: no agents have the power to set prices, the prices are formed through a negotiation process. Finally, the third is the search cost: agents discount their payoff if they do not find a proper trading partner or fail to negotiate. The possibility of not finding a proper trading partner is an important additional search cost in our model. 

It is well known that such complex models are often intractable. However, by  considering large markets and  adopting a mean-field approach, we show that this type of  model becomes tractable and exhibits many interesting properties. In particular, following \cite{NguyenEC2012}, we consider a non-cooperative bargaining game in a finite network, and study the agents' behavior in the limit as the  population at each node of the network increases. We introduce a notion  of a {\it limit stationary  equilibrium}, and show that it always exists.  We then use this concept to investigate the efficiency of the market, how bargaining with middlemen cause endogenous delay in equilibrium, and how network structure influences competition among middlemen and the share of surplus between producers and consumers.  To illustrate these new insights,  next we describe three properties exhibited at equilibrium in our model that are qualitatively different from the predictions in  the existing literature.

\subsection*{Competition among middlemen}
\begin{figure}[htbp]
\centering
\includegraphics[width=1.72in]{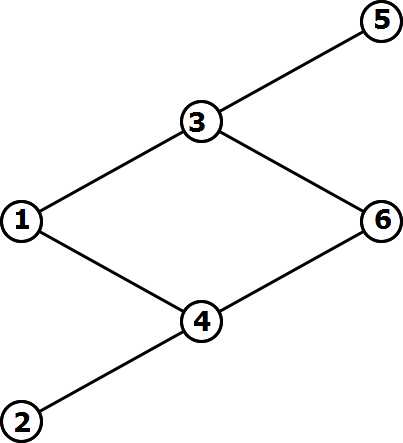}
\caption{An example trading network with producers (1,2);  consumers (5,6) and  middlemen (3,4).}
\label{fig:network0}
\end{figure}

First, to illustrate how network structure influences the competition among middlemen, consider the trading network shown in Figure~\ref{fig:network0}. For simplicity, we assume each producer produces a unit of an indivisible good, and  each consumer desires one item and obtains a  value normalized to 1 dollar upon consuming it. The links in the network represent which pairs of agents can trade and are assumed to be directional links going from left to right. In our model, only pairs of agents that are connected can meet  and bargain.  Unlike a static model like \cite{blume2009trading}, we assume\footnote{A precise description, with a more general model, is given in Section~\ref{model}.} agents meet and trade over a infinite, discrete time horizon, where each agent discounts their payoffs by a factor $0<\delta<1$. 


In static models without search frictions like \cite{blume2009trading}, the middleman have all the bargaining power and they simultaneously suggest prices to the producers and consumers for trade to occur. 
 The producers and consumers then use these prices to determine the middlemen they would like to use for the trade. The middlemen never hold the good, and so need not consider the consequences of any future competition between each other. Such models predict that  the ``bargaining power'' of middlemen at nodes 3 and 4 in this network should be the same, because the roles of agents at these nodes can be interchanged if we change the role of  producers and consumers. In our model, because of search friction, middlemen do not meet both producers and consumers at the same time and so cannot simultaneously propose prices to 
each of them. Only after buying a good from producers, can the middlemen resell it to consumers. Assuming this type of market structure significantly changes the outcome.  In the network in Figure~\ref{fig:network0}, for example, our results show that there is no symmetry between nodes 3 and 4. In particular,  
the bargaining power of middlemen at node $3$ is higher than that of a middlemen at node $4$.
Middleman $3$ has a competitive advantage on the consumer side: it has access to more consumers than middleman $4$. Thus, when holding a good, middlemen $3$ can find a consumer  in an easier and faster manner than $4$. This has an important effect for trade in previous rounds between producers and middlemen. Here, even though $4$ has  access to more producers  (1 and 2), after buying the good from these producers, middleman 4 is aware that he needs to compete and cannot get as high a surplus as  middleman 3, resulting in him not being able to offer good prices to producers 1 and 2.  In other words, competition on the consumers' side has an influence back along the trading network to the competition  on the producers' side, and this disadvantages middleman 4. See Theorem~\ref{coroll0} for a precise statement.

\subsection*{Endogenous Delay}

\begin{figure}[htbp]
\centering
\includegraphics[width=2.5in]{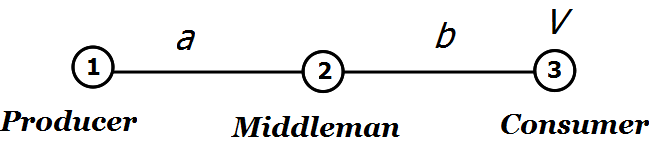}
\caption{A network exhibiting endogenous delay.}
\label{fig:2}
\end{figure}

The second distinguishing property  of our model is \emph{endogenous delay} in trade due to the sunk cost problem. Consider the network shown in Figure~\ref{fig:2}. The producer has to trade with one middleman in order for the good to reach the consumer. Assume that the consumer has a value of $V$ units for consuming the good, and trade on the two links incurs transaction costs of  $a$ and $b$ units, respectively. 

In an efficient market,  if $V>a+b$, so that trade is beneficial,  producers would trade with middlemen and middlemen would trade with consumers whenever these agents meet one another. However, in our model, after buying a good from the producer, middlemen needs to bargain with a consumer to resell the good. At this point, the transaction cost $a$ is sunk and is irrelevant in the negotiation. \cite{WongWright2011} consider a model for such a setting, where each node represents a single agent. In such a model, the expected payoff of the middlemen from the re-sale might not be enough to recover the sunk cost, leading to market failure\footnote{Mathematically, this happens when $V<(1+\alpha) a+b$ for a positive $\alpha$ that depends on other parameters of the model.}. In our model  each node consists of a  large population of agents and the ``bargaining power'' of a middleman agent at location $2$ compared with a consumer agent at location $3$ depends on the competition with  other middlemen that are also trying to sell\footnote{In our model, we assume each middlemen can hold at most one item at a time, thus  middlemen that are holding an item need to sell before buying again.}. In particular,  if the fraction of  middlemen that are selling is small, then when negotiating with consumers, they obtain a higher payoff, which will overcome the sunk cost problem of  trading  with $1$. However,  to maintain the small fraction of middlemen looking to sell, the rate of trades between 1 and 2 needs to be smaller than the rate between 2 and 3. This then implies that when producers and middlemen meet, they do not  trade with probability 1. This can only be rationalized  if the surplus of trade is the same as the producers' outside option, which we normalize to be 0. In other words, in this case the  producers  are indifferent between trading and not trading. When two agents meet, even though they  can potentially trade, if they only enact a successful negotiation with a probability $p \in (0,1)$, we interpret this as endogenous delay.  This result is formally stated in Theorem~\ref{theo:twohops} in Section~\ref{sec:delay}.

\subsection*{Contrast with Double Marginalization}
Lastly, we contrast  the outcome of our model with the classical theory of double marginalization, see for example~\cite{Lerner1934} and \cite{Tirole}.
Double marginalization appears in a similar market structure like ours, where producers sell the good to middlemen and middlemen continue to sell to the consumers downstream.  The market protocols in these environments, however are different from  our model. Namely,  in the double marginalization literature, it is assumed that when selling the good producers and middlemen have total market power, and charge 
a  monopoly price to their downstream market.  As a consequence, middlemen earn a non-negligible profit, consumers pay higher prices and producers have lower profits.

On the other hand, here we assume no agents have a monopoly market power. The prices are formed through a negotiation process. Furthermore, we assume middlemen are long lived, while producers and consumers exit the game after trading\footnote{The assumption of long-lived middlemen and short-lived producers and consumers is also made in \cite{RubinsteinWolinsky1987} and \cite{WongWright2011}.}.
 This captures  the contrast between different type of agents: middlemen often stay in the market for  a long period, while producers and consumers have  limited supply and demand for a certain  good and do not participate in the market after getting rid of the supply or having satisfied the demand.  This assumption captures  many realistic markets  among small producers and consumers, who are faced with search problems and need to trade through middlemen; examples of  such markets include the following: agricultural markets with farmers, consumers and grocery stores; e-commerce markets with sellers, buyers and entities like Ebay or Amazon; and financial markets involving investors, borrowers and banks.

We will show that these assumptions have a fundamental impact on the price formation in the economy. In particular, as the discount rate goes to 1, so that it does not cost agents to wait, the total equilibrium payoff of producers and consumer approaches the total value of trade. What this means is that as agents are more patient, middlemen earn a negligible fee per transaction.  The intuition is that in our model, we assume middlemen stay in the game forever, but do not consume. They instead earn money by flipping the good. Thus, middlemen have an incentive to  buy and sell the good relatively fast. On the other hand as  the discount rate goes to 1, producers and consumer can engage in costless search and bargaining.  This brings down the intermediary fee, and helps the market work more efficiently\footnote{This however does not imply that the total aggregate payoff of middlemen approaches 0. Their  limit payoff is $\sum_{k=1}^\infty{\delta^k}\cdot \mathrm{fee}$ can be positive as $\delta$ approaches 1 and $\mathrm{fee}$ approaches 0.}.  Theorem~\ref{corol1} states this finding rigorously.

 Naturally, when the discount rate is not close to 1, then the above property does not hold. More generally, how the agents share the trade surplus depends on a complex combination of  the network structure and the discount rate. We will rigorously study this question in the rest of our paper.

\subsection{Related Work}

As discussed previously, one line of work that this paper draws upon originates 
from \cite{RubinsteinWolinsky1987} who give a model for search frictions in decentralized 
trade involving middlemen. The setting in \cite{RubinsteinWolinsky1987}  can be viewed
in terms of our model as a simple three node network, with the nodes corresponding to 
producers, consumers and middelmen, respectively; trade may occur either directly 
between a producer and consumer or via a middlemen. 
As in our model, the market evolves in a sequence of periods, where in each period agents are 
matched and discount future profits. However, instead of considering a strategic bargaining model 
as we do, \cite{RubinsteinWolinsky1987} consider a model in which if trade is profitable, it occurs 
with the net surplus being split between the agents. The equilibrium of this market is studied under 
a steady-state assumption. This is similar to the limit-stationary equilibrium that we consider, 
except here we show that such a stationary equilibrium emerges naturally in a limiting sense.

The articles \cite{WongWright2011} and \cite{NguyenEC2012} extend this type of model to line networks, i.e. 
networks consisting of a sequence of nodes $v_1,v_2,\ldots,v_m$ in which trade occurs only between nodes 
$v_n$ and $v_{n+1}$, for $n=1,\ldots,m-1$ (here, $v_1$ is a producer, $v_m$ a consumer and the remaining nodes are middlemen). \cite{WongWright2011} consider a more extensive 
bargaining model and study both a model with one agent per node and a model with many agents per node, under a steady-state assumption similar to \cite{RubinsteinWolinsky1987}.
\cite{NguyenEC2012} considers a similar bargaining model as in this paper but does 
not model search friction in the same way. Specifically, in \cite{NguyenEC2012} the matching process proceeds by first selecting one agent to be a proposer. The proposer is then always able to find a feasible trading partner if one exists (i.e. if the proposer has a good, it is able to find either a consumer or middleman without the good to trade with).  In contrast, in the matching model considered here, a proposing agent may be matched with another agent with whom trade is infeasible, even if a feasible trading partner exists. This increases the search costs and has important consequences. Namely, in \cite{NguyenEC2012}, it is shown that a limit stationary equilibrium might not exist, while here we show that one always does. 

The \cite{RubinsteinWolinsky1987} paper and the aforementioned references focus on the role of middlemen in reducing 
search frictions (see also \cite{yavacs1994middlemen}).  Other works have considered a
middleman's role in mitigating information frictions, including
\cite{biglaiser1993middlemen,li1998middlemen}; such considerations are not addressed here.

Other related models of decentralized trade in networks include \cite{Manea2011bargaining}, which considers distributed bargaining in a network consisting of only producers and consumers as well 
as earlier work including \cite{Rubinstein-Wolinsky, binmore1988matching, gale1987limit}, which 
consider decentralized bargaining between producer and consumers, where any consumer 
can potentially trade with any producer without involving any middlemen.

The other line of work this paper draws on is work such as \cite{blume2009trading}
that focuses on general trading networks and seeks to understand how network structure
 impacts the division of the gains from trade. As in this paper, \cite{blume2009trading} consider general networks with the restriction that all trade must go through a middleman and middlemen do not 
trade with each other. Here, we also do not consider networks in which middlemen trade with 
each other, but we do allow for trade routes that do not involve any middlemen. As noted previously, in \cite{blume2009trading} middlemen can simultaneously announce prices to buyers and sellers. The full information Nash equilibrium of the resulting pricing game is characterized. Somewhat related equilibria questions have been studied in the context of supply chains (see e.g.~\cite{nagurney2002supply}), 
where in this case, producers are able to produce and ship multiple units of a product to middlemen and consumers.

On the technical side, our solution concept of limit stationary equilibria is closely related to 
work on mean field equilibrium\footnote{In this line of work, the convergence of finite-player games to mean-field equilibria is rigorously analyzed so that any spurious mean-field equilibria can be rejected. See~\cite{Gomes:2010fk, Adlakha:2010uq}, for example.} for dynamic games (see e.g.~\cite{graham1994chaos, lasry2007mean, gueant2011mean, Benaim:2011fk}). As in our analysis, the theme in this work is 
characterizing a notion of equilibria for a ``large market," in which users make 
decisions based on a steady-state view of the market, where this steady-state view 
is asymptotically consistent with the user's actions. In most of the mean field literature, all users are statistically identical, while in our model each user has a fixed type depending  on his location in the trading network. This notion is similar to work in \cite{huang2010nce}, which considers a
mean field limit for the  control for linear quadratic Gaussian systems where the interaction between users depends on their ``locality."

The remainder of the paper is organized as follows.  Section 2  introduces the baseline non-cooperative bargaining model. Section 3 discusses the solution concept of limit stationary equilibrium. Sections 4 and 5 uses this equilibrium to provide comparative analysis of several networks.  Section 6 concludes.

\section{The Model}\label{model}
In this section we introduce the model that we will use. We start by defining the concept of a trading network.

\subsubsection*{Trading Network:} 
\begin{figure}[htbp]
\centering
\includegraphics[width=3.1in]{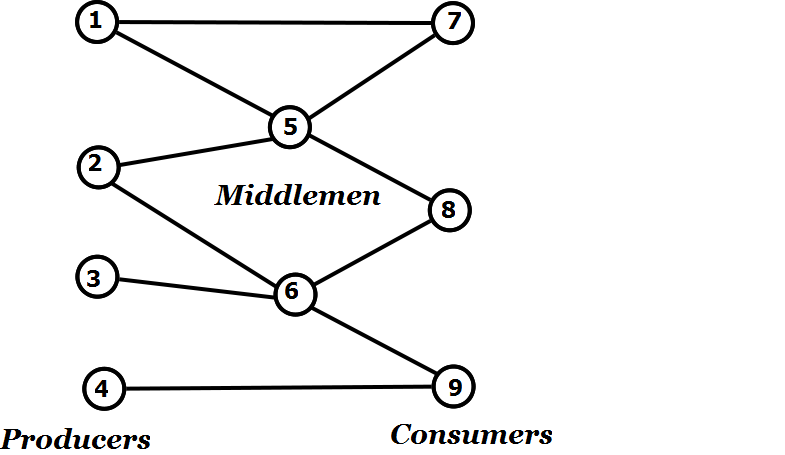}
\caption{A network among producers, consumers and middlemen.}
\label{fig:1}
\end{figure}

We consider a group of producers, consumers and middlemen interconnected by an underlying trading network, which is modeled as a directed graph, $\mathcal{G}=(\mathcal{V},\mathcal{E})$ (see Figure~\ref{fig:1}). Each node $i\in \mathcal V$ represents a population of $N_i$ agents, all of which are either consumers, producers or middlemen.
 Hence, we can partition the set of vertices into the following three disjoint sets: a set of producers denoted by $\mathcal{P}$, a set of middlemen denoted by $\mathcal{M}$, and a set of consumers denoted by $\mathcal{C}$. An agent from the population at  a node $i$ will sometime be referred to as a type $i$ agent. Trade occurs over directed edges, i.e., a directed edge $(i,j) \in \mathcal {E}$ indicates that a type $i$ agent  can potentially directly trade with any type $j$ agent with the good going from $i$ to $j$ as a result of the trade. With a slight abuse of terminology, we often refer to two such agents as being connected by the edge $(i,j)$.
For a consumer to acquire a good from a producer, there must be a (directed) path from the consumer to the producer. If this path has length 1, then the two can directly trade, otherwise they must rely on middlemen to facilitate the trade.
For simplicity, we consider networks in which any path between a consumer and producer contains at most one middleman, i.e
all such paths are either length 1 or 2. An example of such a network is shown in Figure~\ref{fig:1}. With this assumption, the set of directed edges, $\mathcal{E}$ can also be partitioned into three disjoint sets: those that directly connect producers to consumers (denoted by $\mathcal{E}_1$), those that connect producers to middlemen (denoted by $\mathcal{E}_2$), and those that connect middlemen to consumers (denoted by $\mathcal{E}_3$).

We assume that there is one type of indivisible good in this economy\footnote{The analysis easily extends to a finite number of distinguishable goods.}. All producers produce identical goods and all consumers want to acquire these goods.  The value that each consumer of type $c\in \mathcal{C}$ gets from an item is 
$V_c\ge 0$. In every period each agent can hold at most one unit of the good (an item)\footnote{Again the analysis easily extends to allowing agents to hold a finite number of goods, \emph{albeit} at the cost of more notation and laborious book-keeping.}. Thus, in every time period,  a middleman either has an item or does not have one. Hence, if there is a directed edge from node $i$ to node $j$, a specific agent of type $i$ can only trade with an agent of type $j$ if the type $i$ agent has a copy of the good and the type $j$ agent does not; we refer to such a pair of agents as {\it feasible trading partners}. Note that producers are assumed to always have a good available to trade and consumers are always willing to purchase a good. So, for example, any two agents connected by an edge in the set $\mathcal{E}_1$ are always feasible trading partners. For every edge $(i,j)\in \mathcal{E}$, we associate a non-negative transaction cost $C_{ij}\ge 0$; this cost is incurred when trade occurs between an agent at node $i$ and one at node $j$.

Next we discuss the bargaining process that determines the trading patterns for how goods move through the network.

\subsubsection*{The Bargaining  Process:} 
We consider  an infinite horizon, discrete time repeated bargaining game, where agents discount their payoff at rate $0<\delta<1$\footnote{The model can be extended to allow for heterogeneous discount rates.}.
Each period has multiple steps and is described as follows.

{\bf Step 1.} One among all pairs of directly connected nodes $(i,j)\in \mathcal{E}$ is selected at random with a predetermined probability distribution $\pi_{ij})$ on the set of edges $\mathcal{E}$ and one node from each of the corresponding populations is selected uniformly at random. One of these agents is further selected to be a proposer, again chosen at random\footnote{The model easily generalizes to allow for different but fixed probabilities of picking each end-point of a link as the proposer of a trade.}.

{\bf Step 2.} If the agents are not feasible trading partners, then the  game moves to the next period and restarts at step 1. Recall that this will occur if neither agent has the good or if both have the good.

{\bf Step 3.} The proposer makes a take-it-or-leave it offer of a price at which he is willing to trade.  If the trading partner refuses, the game moves to the next period. Otherwise, the two agents trade: one agent gives the item  to and receives the money from the other, and the proposer pays for the transaction cost $C_{ij}$\footnote{Actually, it does not matter who pays for this transaction cost, because, in equilibrium, the transaction cost is reflected in the proposed price.}.  If a consumer or producer participates in a trade, they exit the game and are replaced by a clone. On the other hand, middlemen are long lived and do not produce nor consume; they earn money by flipping the good.

{\bf Step 4.} The game moves to the next period, which starts from Step~1.

The game is denoted by $\Gamma(\mathcal{G},\vec{C},\vec{V},\vec{N},{\delta})$, where $\vec{C}$ denotes the vector of links costs, $\vec{V}$ denotes the vector of consumer valuations and $\vec{N}$ denotes the vector of population sizes at each node. Sometimes, we will simply refer to this game as $\Gamma$. 

{\it Remarks:} We assume middlemen are long lived, on the other hand, producers and consumers exit the game after trading. This captures an extreme contrast between different type of agents: middlemen often stay in the market for  a long period, while producers and consumers have limited supply and demand for a certain  good. 
The assumption that we make about the replacement of producers and consumers capture a steady state of an extended  economy, where there are incoming flows of producers and consumers  at certain rates.  Here, to focus on the solution in a steady state, we assume these incoming rates are equal to the rates at which these agents successfully trade. This can be made  endogenous as in \cite{gale1987limit}. However, in a fully 
endogenous model, a characterization of equilibrium is difficult.  Here, we shortcut this problem by assuming the economy is already in a steady state.

In Steps 1 and 2, it is possible that trade is not possible between agents identified at the ends of the chosen link. This leads to additional search friction as it results in a loss of trading opportunity for these two agents. By appropriately choosing the distribution $\pi$ and the choice of proposing agent when trade takes place, we can equivalently view the dynamics from the perspective of the nodes such that the agents are picked independently (following some distribution) to be proposers and depending on the state of the agent (i.e., if the agent possesses the good or not), one among the appropriate edges is chosen following a distribution. Note that even from the node perspective, there is a possibility that the proposing agent might pick an edge along which no trade is possible owing to the picked agent having the same state as the proposing agent; once again, this leads to additional search friction. We prefer to model the dynamics from the perspective of edges as it is more general and fully subsumes the node perspective.



Since we are interested in analyzing the bargaining process defined earlier, as the number of agents in each location increases without bound, we proceed to precisely define the effect of increasing the population size.

\subsubsection*{Replicated  Economy:}
Given the bargaining game $\Gamma(\mathcal{G},\vec{C}, \vec{V}, \vec{N},{\delta})$, the game's $k$th replication is defined as  a game of the same structure except the population size  is increased by a factor of $k$ at each node, and the time gap between consecutive periods is reduced by a factor of $T_k$. Formally, this is defined as follows:


\begin{definition}\it
Given the game $\Gamma(\mathcal{G},\vec{C}, \vec{V},\vec{N}, {\delta})$ and $k, T_k\in \mathbb{N}_+$, let 
$\delta'= \delta^{1/T_k}$. Then the $(k,T_k)$-replication of $\Gamma$, denoted by 
$\Gamma^k_{T_k}(\mathcal{G},\vec{C},\vec{V}, \vec{N}, {\delta})$ is defined as
$\Gamma(\mathcal{G},\vec{C},\vec{V},k\vec{N}, \delta')$.
\end{definition}

{\it Remark:} The scaling of the discount rate $\delta$ in this definition is commonly used in the study of  dynamical systems.
It is clear that without changing $\delta$, in the replicated economy each agent will need to wait for a longer and longer time to get selected, and thus his  pay-off  approaches $0$.  If initially each period takes one unit of time, then note that changing the discount rate to $\delta'= \delta^{1/T_k}$ is mathematically equivalent to  changing  the time gap between periods to become  $1/T_k$ time units and keeping the discount rate fixed. Hence, for example, if we choose $T_k=c\cdot k$, it  means we keep the rate that each agent sees trading opportunities on the same order as in the original finite game. On the other hand  $T_k>>k$ models a setting in which the rate at which agents trade is increasing. In this paper, for simplicity, we  will  focus on the case $T_k=k$. Other choices of $T_k$ do not affect our results, qualitatively.

\section{Solution Concept And Existence Of Equilibrium}\label{equilibrium}
Next we turn to the solution concept considered in this paper, that  we call a {\it limit stationary equilibrium}.  To define this equilibrium, we follow \cite{NguyenEC2012} and  consider the limit of finite agent games as the population increases. In particular, in each game with finite population, we consider a semi-stationary equilibrium in which  each agent  believes that the economy is already in a {\em steady state} and behaves according to a {\em stationary  strategy} profile.  This is certainly not enough. To ``close the loop,''  a limit stationary equilibrium is defined as a limit of semi-stationary equilibria, whose dynamics  converge to the presumed state.

 
To be more precise,  we have the following definitions.


\begin{definition}\it
The {\it state of the economy} is a vector $\vec{\mu}\in [0,1]^{|\mathcal{M}|}$, where $\mu_m$ denotes the fraction of middlemen at node $m$ that hold an item.
\end{definition}

\begin{definition}\label{strategy} \it
A strategy profile (possibly mixed strategy) is called a {\it stationary  strategy}  if  it only depends on an agent's  identity, his state (owning or not owning and item) and the play of the game (which  agent he is bargaining with, who the proposer is and what is  proposed). More precisely,  suppose  that agent $i$ and agent $j$ are selected to bargain, and assume $i$ owns an item, $j$ does not, furthermore   $i$ is the proposer. In this case,  a stationary  strategy of agent $i$ is a distribution of proposed prices to agent $j$ and a stationary  strategy of agent $j$ is a probability of accepting  the offer.

In the rest of the paper, given a stationary strategy, for each link $(i,j)\in \mathcal E$ we let $\lambda_{ij}$ denote the  conditional probability that $i$ and $j$ trade when they are matched and trade is feasible, that is  $i$ owns an item and $j$ does not.
\end{definition}

In the following, we first start with the definition of a semi stationary equilibrium in a finite economy, we then use this concept to define limit-stationary equilibria as the size of the economy increases.

\subsection{Semi-stationary equilibria}
Informally, given a  finite game and a state $\vec{\mu}$, a stationary  strategy profile is a semi-stationary equilibrium of the game with respect to  $\vec{\mu}$, if under the hypothesis that agents believe the state of the economy is always  $\vec{\mu}$,  no agent can strictly improve his payoffs by  changing his  strategy\footnote{This stationary belief discounts the impact of certain strategic behaviors that will vanish owing to competition between agents at each node as the population size increases, when determining the incentive constraints. In particular, consider the case where there are exactly two consumers each of a different type connected to a single middleman with one consumer being better than the other, both in terms of a higher value for the good and a lower transaction cost. Then, the better consumer could refuse to trade unless he gets a higher payoff by being offered the same price as the other consumer.}.

To define this concept more precisely, we need to introduce the  expected pay-offs  of agent $i$  depending on whether he possesses or does not possess a good, which we denote by $u_0(i)$ and $u_1(i)$, respectively.  
Notice that because of the assumption that producers and consumers exit the market after a successful trade, we have $u_0(p)=0$  for all $p\in \mathcal{P}$ and  $u_1(c)=V_c$ for all $c\in \mathcal{C}$. Furthermore, we assume that all agents believe that the state of the economy is captured by $\vec{\mu}$.
For the present, we will assume that $\vec{\mu}$ is given. After deriving the incentive conditions depending on $\vec{\mu}$, we will discuss the second type of conditions that give  $\vec{\mu}$ endogenously.

The basic structure of the incentive constraints can be captured by the following argument. Assume two agents $i$ and $j$ meet, where $i$ holds the good and $j$ wants it. Also assume that $i$ is the proposer.   If the trade is successfully completed, then $j$  possesses the item, thus agent $i$ will demand from  agent $j$ the difference of the payoffs between the states before and after the trade (discounted by $\delta$). Note that the state of $i$ also changes, and therefore, if trade is successfully completed, then $i$'s payoff is
\begin{align*}
\delta u_0(i) + \delta \big( u_1(j)-u_0(j) \big) - C_{ij}.
\end{align*}
However, agent $i$ has the option of not proposing a trade (or proposing something that will necessarily be rejected by the other party) and earn a payoff of $\delta u_1(i)$. Thus, in this situation, the continuation payoff of agent $i$ is 
\begin{align*}
\max\{\delta u_1(i), \delta u_0(i) + \delta \big( u_1(j)-u_0(j) \big) - C_{ij}\}.
\end{align*}

 For ease of exposition define the difference between the two terms in this maximization to be 
\begin{align}
z_{ij} := \delta \Big(u_1(j) - u_0(j) - \big(u_1(i)-u_0(i) \big)\Big)-C_{ij}. \label{eq:zab}
\end{align}
Thus,  the continuation payoff of agent $i$ when he is proposing to $j$ is
$$
\delta u_1(i) + \max\{z_{ij},0\}.
$$
From this we also obtain  the following conditions on the dynamics of trade:
\begin{enumerate}
\item If $z_{ij}<0$,  then agent $i$ will never sell an item to agent $j$ and will wait for a future trade opportunity;
\item If $z_{ij}>0$,  then agent $i$ will sell the item to agent $j$ with probability one whenever they are matched;\footnote{Note that when $z_{ij}>0$ in equilibrium, it has to be the case that if agent $i$ proposes to trade agent $j$ will agree to the trade. This is  because if $j$ only agrees with a probability $0<p<1$, $i$ can improve his payoff  by  decreasing the proposing price by a small $\epsilon>0$. However, for any such  $\epsilon>0$, $i$ again has a better deviation by decreasing the proposing price by a smaller amount, say $\epsilon/2$.} and finally
\item If $z_{ij}=0$, then agent $i$  is indifferent between  selling and waiting, thus,  the trade can occur with some probability $\lambda_{ij} \in [0, 1]$. Conversely,  if trade between agents $i$ and $j$ occurs with probability $0< \lambda_{ij} < 1$, then we must have $z_{ij}=0$.
\end{enumerate}

Similarly, assume now that instead of $i$, agent $j$ is the proposer, then the continuation payoff of $j$ in this case is $ \delta u_0(j) + \max\{z_{ij},0\}.$ Furthermore, the same conditions concerning the dynamic of trade between $i$ and $j$, which depends on $z_{ij}$ hold as above, but with the roles of $i$ and $j$ interchanged.

These conditions can be delineated for the general network model introduced in the previous section by considering each type of agent, the state of the agent in terms of holding a good or not, and the probability that he is selected as a proposer. In our model, we have three types of agents: producers, consumers and middlemen. Middlemen are active in the game regardless of having or not having an item.  Thus, we will need four types of equations expressing the expected payoff of these agents given their states.

We consider these conditions in detail for the case of  producers; the rest follows in similar fashion using the logic outlined earlier. For each producer  of type $p\in\mathcal{P}$ who has an item to sell in each period, an agent $p$'s continuation payoff depends on  which type of link is selected,  the pair of agents that are selected to trade, and whether $p$ is selected as the proposer. Thus, agent $p$'s  expected continuation payoff is  
\begin{align}\label{eq:producer}
\sum_{c: (p,c) \in \mathcal{E}_1}  \frac{\pi_{pc}}{2 N_p } (\delta u_1(p)+\max\{z_{pc},0 \}) + \sum_{m: (p,m) \in \mathcal{E}_2}  \frac{\pi_{pm}}{2 N_p  } (1-\mu_m) (\delta u_1(p)+ \max\{z_{pm},0 \}) + \\ \nonumber
 +\big(1-\sum_{c: (p,c) \in \mathcal{E}_1}  \frac{\pi_{pc}}{2 N_p }-\sum_{m: (p,m) \in \mathcal{E}_2}  \frac{\pi_{pm}}{2 N_p  } (1-\mu_m)\big )\delta u_1(p)
\end{align}
Here, $z_{pc}$ and $z_{pm}$ are defined as in (\ref{eq:zab}). The first term of \eqref{eq:producer} represents the case where $p$ is the proposer to a consumer $c$. The second term represents $p$ proposing  to a middlemen $m$, who currently does not own a good. Finally, the last term describes the case where $p$ is not a proposer. Here, recall that $N_p$ is  the size of population at  node $p$,  and thus,   $\frac{\pi_{pc}}{2 N_p }$, is the probability that the specific agent of type $p$ is the proposer\footnote{It is in this calculation that the generalization to different probabilities for the choice of a proposer can be added.} for a consumer $c$. On the other hand, because only a fraction of middlemen are looking to buy, for  every middlemen node $m\in\mathcal{M}$, $\frac{\pi_{pm}}{2N_p}(1-\mu_m)$ is the probability that  $p$ is matched with $m$, $m$ does not  hold a good and $p$ is the proposer. One can interpret this as a form of {\it search friction}, that is the probability that a producer can find a trade-able middlemen depends on the state of the economy, which, in turn, impacts the transaction dynamics between the producer and the middleman. 

Now, because $\vec{u}$ are assumed to be values of a stationary equilibrium, $u_1(p)$ needs to equal the expression in \eqref{eq:producer}. After some algebraic  manipulation, this is equivalent to 
\begin{align}\label{eq:sell1f}
u_1(p) = \sum_{c: (p,c) \in \mathcal{E}_1}  \frac{\pi_{pc}}{2 N_p (1-\delta) } \max\{z_{pc},0 \} + \sum_{m: (p,m) \in \mathcal{E}_2}  \frac{\pi_{pm}}{2 N_p (1-\delta) } (1-\mu_m) \max\{z_{pm},0 \}.
\end{align}

Similarly, for  the two type of middlemen (either owning an item or not) and  the consumers, we have the following set of equations:
\begin{align}
\forall m \in \mathcal{M} \quad u_0(m) & = \sum_{p: (p,m) \in \mathcal{E}_2} \frac{\pi_{pm}}{2 N_m (1-\delta)} \max\{z_{pm},0\}, \label{eq:limj0f} \\
\forall m \in \mathcal{M} \quad u_1(m) & = \sum _{c: (m,c) \in \mathcal{E}_3} \frac{\pi_{mc}}{2 N_m (1-\delta) } \max\{z_{mc},0 \}, \label{eq:limj1f}\\
\forall c \in \mathcal{C} \quad u_0(c) & = \sum_{p: (p,c) \in \mathcal{E}_1}  \frac{\pi_{pc}}{2 N_c (1-\delta) }\max\{z_{pc},0 \} 
+ \sum _{m: (m,c) \in \mathcal{E}_3}  \frac{\pi_{mc}}{2 N_c (1-\delta) } \mu_m \max\{z_{mc},0 \}, 
\label{eq:limk0f}
\end{align}
where  $z_{pm}, z_{mc}, z_{pc}$ are defined as
\begin{equation}\label{eq:zabf}
z_{ij}= \delta\Big(u_1(j) - u_0(j) - \big(u_1(i)-u_0(i) \big)\Big)-C_{ij} \;\;\; \forall (i,j) \in \mathcal{E}_1 \cup \mathcal{E}_2 \cup \mathcal{E}_3.
\end{equation}
Once again, the state $\vec{\mu}$ appears in the incentive equations above owing to the particular search model that we consider. 
As mentioned earlier, since producer and consumers exit the game after trading successfully,  we have 
\begin{align}
\forall p \in \mathcal{P} \quad u_0(p) & = 0, \label{eq:u0if}\\
\forall c \in \mathcal{C} \quad u_1(c) & = V_c. \label{eq:u1kf}
\end{align}

We are now ready to define a semi-stationary equilibrium.
\begin{definition}\it
Given a finite game $\Gamma(\mathcal{G},\vec{C},\vec{V},\vec{N},{\delta})$ and a state $\vec{\mu}$, a stationary strategy profile is a semi-stationary equilibrium with respect to $\vec{\mu}$ if and only if there exists $\vec{u}, \vec{z}$ satisfying \eqref{eq:sell1f}-\eqref{eq:u1kf}, and furthermore, 
\begin{itemize}
\item If $z_{ij}<0$,  then irrespective of who the proposer is, agent $i$ will never sell an item to agent $j$, so that he will wait for a future trade opportunity;
\item If $z_{ij}>0$,  then irrespective of who the proposer is, agent $i$ will sell the item to agent $j$ with probability one whenever they are matched; and finally
\item If $z_{ij}=0$, then proposer of the trade  is indifferent between trading and waiting. Thus,  the trade occurs with some probability $0\le \lambda_{ij} \le 1$. In addition, if trade between agents $i$ and $j$ occurs with probability $0< \lambda_{ij} < 1$, then $z_{ij}=0$.
\end{itemize}
\end{definition}

\subsection{Limit stationary equilibrium}
As the economy gets large, we need to consider the behavior of equations \eqref{eq:sell1f}-\eqref{eq:u1kf}, where $N_i$ is replaced by $k N_i$ and $\delta$ is replaced by $\delta^{1/k}$, as $k$ increases without bound. Note that $$\lim_{k\rightarrow\infty} k (1-\delta^{1/k})  = \ln(1/\delta) \text{ and } \lim_{k\rightarrow\infty} \delta^{1/k}=1.$$

Hence, in the limit, the set of equations \eqref{eq:sell1f}-\eqref{eq:u1kf} yield the following:
\begin{align}
\begin{split}
\forall p \in\mathcal{P}  \quad u_0(p) & = 0;\\
\quad u_1(p)  &= \sum_{c: (p,c) \in \mathcal{E}_1}  \frac{\pi_{pc}}{2 N_p \ln(1/\delta) } \max\{z_{pc},0 \} 
+ \sum_{m: (p,m) \in \mathcal{E}_2}  \frac{\pi_{pm}}{2 N_p \ln(1/\delta) } (1-\mu_m) \max\{z_{pm}, 0 \} ;
\end{split} \label{eq:limi1}\\
\forall m \in \mathcal{M} \quad u_0(m) & = \sum_{p: (p,m) \in \mathcal{E}_2} \frac{\pi_{pm}}{2 N_m \ln(1/\delta)} \max\{z_{pm},0\}; \label{eq:limj0} \\
\forall m \in \mathcal{M} \quad u_1(m) & = \sum _{c: (m,c) \in \mathcal{E}_3} \frac{\pi_{mc}}{2 N_m \ln(1/\delta) } \max\{z_{mc},0 \}; \label{eq:limj1}\\
\begin{split}
\forall c \in \mathcal{C} \quad u_0(c) & = \sum_{p: (p,c) \in \mathcal{E}_1}  \frac{\pi_{pc}}{2 N_c \ln(1/\delta) }\max\{z_{pc},0 \} 
+ \sum _{m: (m,c) \in \mathcal{E}_3}  \frac{\pi_{mc}}{2 N_c \ln(1/\delta) } \mu_m \max\{z_{mc},0 \};\\
\quad u_1(c) & = V_c ; \text{ and}
\end{split}  \label{eq:limk0}\\
 z_{ij} &= \Big(u_1(j) - u_0(j) - \big(u_1(i)-u_0(i) \big)\Big)-C_{ij} \;\;\; \forall (i,j) \in \mathcal{E}_1 \cup \mathcal{E}_2 \cup \mathcal{E}_3.
 \label{eq:limzab}
\end{align}

Using these limiting equations, we now define our solution concept of limit stationary equilibrium. Intuitively, a limit stationary equilibrium is a  stationary strategy   that can be associated with $\vec{u}, \vec{z}$ satisfying \eqref{eq:limi1}-\eqref{eq:limzab} in an similar way to the definition  of semi-stationary equilibrium.  However, here another condition is added to guarantee that the dynamic given by the stationary strategy will actually converge to the presumed state of the economy $\vec{\mu}$.
In particular, recall that  a stationary strategy is given by a set of probabilities $\lambda_{ij}$ for every edge $(i,j) \in \mathcal{E}$, which denotes the probability of  trade occurring among a pair of {\em feasible trading partners} of type $i$ and $j$. If $\lambda_{ij} = 1$, trade always occurs and if $\lambda_{ij} = 0$ it never occurs. Given any such stationary strategy, the resulting dynamics can be modeled by a Markov process, with the state space being the number of agents at each node holding a good.  As the population at each node of the network increases, we want the limit of the stationary distributions of these Markov process to be $\vec{\mu}$. We elaborate on the convergence issue in much greater detail in Section~\ref{sec:mucon}.


More formally, we have the following definition for a limit stationary equilibrium.
 \begin{definition}\it 
Given a finite game $\Gamma(\mathcal{G},\vec{C},\vec{V},\vec{N},{\delta})$  a stationary strategy profile  is a limit  stationary equilibrium if the stationary distribution of the associated Markov process converges to a point-mass\footnote{In a general mean-field analysis setting~\cite{MR1108185}, the final object is typically a product measure over the different types, instead of a point-mass as here. Then, the analysis of \eqref{eq:limi1}-\eqref{eq:limzab} would be carried out by taking expectations over the corresponding product measure.} on the state $\vec{\mu}$, and there exists $\vec{u}, \vec{z}$ satisfying \eqref{eq:limi1}-\eqref{eq:limzab},  moreover 
\begin{itemize}
\item If $z_{ij}<0$,  then irrespective of who the proposer is, agent $i$ will never sell an item to agent $j$, so that he will wait for a future trade opportunity and $\lambda_{ij}=0$;
\item If $z_{ij}>0$,  then irrespective of who the proposer is, agent $i$ will sell the item to agent $j$ with probability one whenever they're matched so that $\lambda_{ij}=1$; and finally
\item If $z_{ij}=0$, then proposer of the trade  is indifferent between trading and waiting. Thus,  the trade occurs with some probability $0\le \lambda_{ij} \le 1$. In addition, if trade between agents $i$ and $j$ occurs with probability $0< \lambda_{ij} < 1$, then $z_{ij}=0$.
\end{itemize}
\end{definition}

\subsection{Existence of a Limit stationary equilibrium}

We next show that a limit stationary equilibrium always exists.  

\begin{theorem}\label{theo:existence}
For a bargaining game, $\Gamma(\mathcal{G},\vec{C},\vec{V},\vec{N},\delta)$, a 
limit stationary equilibrium always exists.
\end{theorem}
\proof{Proof of Theorem~\ref{theo:existence}:}
See Appendix~\ref{sec:proofexist}.

The proof  of this theorem is based on a standard fixed-point theorem argument for the best-response correspondences of a fictitious game that is obtained from the incentive constraints and the trading dynamic. The existence of limit stationary equilibria follows as a consequence of the Markov dynamic resulting from our search model, a point that will be elaborated on in the next section. We note, however, that the limit stationary equilibrium might not be unique. In Section~\ref{networkstruct} we will give a method to compute and check if a stationary  strategy is a valid a limit stationary equilibrium. 

\subsection{Convergence of state of the trading process}\label{sec:mucon}

As mentioned earlier, one of the  key technical lemmas needed to prove the existence of a limit equilibrium is to show that given any stationary strategy profile, the corresponding Markov process in the replicated system will always converge to an unique state, which is also a continuous function of the parameters.  This, in return, allows us to define a continuous mapping between two product  state-payoff spaces, that can be used in a fixed point argument. In this section, we discuss how to define the state of the trading process and how to determine the $\vec{\mu}$ used for defining both a semi-stationary equilibrium and a limit stationary equilibrium. The formulation of the converging state given our particular Markov process will allow us to compute and construct an equilibrium in various networks in the next section\footnote{Readers who are familiar with Markov processes and their mean-field analysis may skip this part.}.



We will prove the existence of $\vec{\mu}$ by analyzing the Markov process that drives the state of the system, for a given set of stationary strategies $\{\lambda_{ij}\}$ for $(i,j)\in\mathcal{E}$. Since the state of middlemen can change with time, the entire system can be represented by a vector-valued random process $\{ X^k_m(t): m\in \mathcal{M}\}_{t=1}^\infty$ where for the $k^{\mathrm{th}}$ replicated system we keep track of the number of agents who have the item at each middleman type $m\in \mathcal{M}$. For mathematical convenience, we will append $\{X^k_p(t): p \in\mathcal{P}\}$ where $X^k_p(t)\equiv k N_p$ and $\{X^k_c(t): c \in\mathcal{C}\}$ where $X^k_c(t) \equiv 0$ for the states of the producers and the consumers, respectively. Since producers exit the game as soon as they sell their good and are replaced by a clone with a good, at any given time any producer always possesses a good. A similar reasoning holds for the consumers never having a good.

For the $k^{\mathrm{th}}$ replication, the state transitions are given as follows for each $m\in \mathcal{M}$
\begin{align*}
X^k_m(t+1) & = 
\begin{cases}
\min(k N_m,X^k_m(t)+1) & \text{w. p. }  \rho^k_{m}(+1) \\
\max(0,X^k_m(t)-1) & \text{w. p. } \rho^k_m(-1)\\
X^k_m(t) & \text{w. p. } 1-\rho^k_m(+1)-\rho^k_m(-1), 
\end{cases}
\end{align*}
where 
\[
\rho^k_{m}(+1) = \left(1-\frac{X^k_m(t)}{kN_m}\right) \sum_{p\in \mathcal{P}: (p,m) \in \mathcal{E}_2} \pi_{pm}\lambda_{pm} 
\]
is the probability that an agent of type $m$ acquires a good in a given period, and 
\[
\rho^k_{m}(-1) = \frac{X^k_m(t)}{kN_m}  \sum_{c\in \mathcal{C}: (m,c) \in \mathcal{E}_3} \pi_{mc} \lambda_{mc} 
\]
is the probability that a type $m$ agent sells a good in a given period. As noted above, the states corresponding to producers in $\mathcal P$ and consumers in $\mathcal C$ are fixed for all time. This shows that we have a Markov process, and it is easily verified that the process is irreducible. 

Since the transition matrix of our Markov process satisfies Lipschitz conditions (see \cite{EthierKurtz2005}), we can analyze the fluid limit that is obtained by scaling time and space, i.e., by considering the process $\{\tilde{X}_v^k(t): v \in \mathcal {V}\}$, where
\begin{align*}
\tilde{X}_v^k(t) := \frac{X_v^k(\lceil k t \rceil)}{k}, \qquad \forall v\in \mathcal{V}, 
\end{align*}
where $\lceil x \rceil$ for real $x$ is the smallest integer greater than $x$.
We will analyze the behavior of the scaled process $\{\tilde{X}_v^k(t): v\in\mathcal{V}\}_{t\in \mathbb{R}_+}$ when $k$ increases without bound. Note that this is the exact scaling considered by the replicated systems discussed earlier. We then have the following result.

\begin{lemma}\label{theo:converge}
Given a set of probabilities for trade $\{\lambda_{ij}, (i,j)\in \mathcal{E}\}$, the stationary distribution of the trading dynamic process described above converges to a point-mass on a unique state $\vec{\mu}$, which is the unique solution of 
\begin{align}\label{eq:balance}
\sum_{c\in \mathcal{C}} \pi_{mc}\mu_m\left(1-\mu_c\right)\lambda_{mc} =  \sum_{p\in\mathcal{P}} \pi_{pm}\mu_p\left(1-\mu_m\right)\lambda_{pm} \quad \forall m\in\mathcal{M},
\end{align}
and which is given by
\begin{align}
\forall p\in\mathcal{P},\quad \mu_p & = 1; \qquad \qquad
\forall c\in\mathcal{C},\quad \mu_c  = 0; \label{eq:dynamic1} \\
\forall m\in \mathcal{M}, \quad \mu_m & = \frac{\sum_{p\in \mathcal{P}: (p,m) \in \mathcal{E}_2} \pi_{pm} \lambda_{pm} }{\sum_{p\in \mathcal{P}: (p,m) \in \mathcal{E}_2} \pi_{pm} \lambda_{pm} +  \sum_{c\in \mathcal{C}: (m,c) \in \mathcal{E}_3}\pi_{mc} \lambda_{mc}}. \label{eq:dynamic2}
\end{align}
\end{lemma}
\proof{Proof of Lemma~\ref{theo:converge}:}
See Appendix~\ref{sec:proofMarkov} for details.

For any fixed $k$, the Markov process $\{\tilde{X}_v^k(t): v \in \mathcal {V}\}$ is irreducible and has finite-states. Therefore, it has a unique stationary distribution. Note that Lemma~\ref{theo:converge} asserts that these stationary distributions converge to a point-mass\footnote{As mentioned earlier, in a general mean-field analysis setting \cite{MR1108185}, the final object need not be a point-mass but a product measure over the different types.} which is determined by stationary behavior of the limiting process obtained as $m$ increases with bound. In effect, the result above justifies an exchange of the order of limits, $k$ first and time later versus time first and $k$ later. It is both the convergence to a point-mass and this exchange of limits that justifies our definition of semi-stationary equilibria and limit stationary equilibria.

From Lemma~\ref{theo:converge}, $\mu_m$ for middlemen of type $m$ is the stationary fraction of agents at node  $m\in\mathcal{M}$  that hold the good.  
On the other hand, for the producers we have $\mu_p\equiv 1$ for all $p\in\mathcal{P}$ and the consumers we have $\mu_c\equiv 0$ for all $c\in\mathcal{C}$. The given values $\{\mu_m: m\in\mathcal{M}\}$ satisfy \eqref{eq:balance}, which can be interpreted as a balance condition: for every node $m\in\mathcal{M}$,  in every period, the probability that the amount of goods held at node $m$ increases by one or decreases by one should be equal.  
Looking at this interpretation of \eqref{eq:balance} more closely, each term on the left-hand-side for a given $c$, $ \pi_{mc}\mu_m\left(1-\mu_c\right)\lambda_{mc}$ is the probability that trade occurs from $m$ to $c$, which requires that link $(m,c)$ is selected (with probability $\pi_{mc}$), that $m$ and $c$ are feasible trading partners (with probability $\mu_m (1-\mu_c)$ so that $m$ has the good and $c$ needs it) and that trade occurs (with probability $\lambda_{mc}$). Similarly, each term on the right-hand-side for a given $p$ can be interpreted as the probability of trade from $p$ to $m$. While the use of $\{\mu_p: p\in\mathcal{P}\}$ and $\{\mu_c: c\in\mathcal{C}\}$ above is for mathematical convenience, in more general networks (for future work) where we allow middlemen to trade with each other, expressions similar to \eqref{eq:balance} will hold as the balance condition for every middlemen type where the terms will involve the state of other middlemen as well.

\section{Competion Among Middlemen}\label{networkstruct}
In this section, we focus on the equilibrium of our model for the network illustrated in Figure~\ref{fig:network2}.  For simplicity, we assume all the transaction costs are 0 and each consumer's valuation  of the good is 1. We will analyze the equilibrium as  $\delta$ takes values between 0 and 1. Note that $\mathcal{P}=\{1, 2\}$, $\mathcal{M}=\{3, 4\}$ and $\mathcal{C}=\{5, 6\}$.


As discussed in the introduction, because producers at node $1$  have access to more middlemen than at node $2$, producers at $1$ have better bargaining power than producers at $2$. Similarly, consumers  at node $6$ are in a better position than consumers at $5$. However, a comparison between the two middlemen position $3$ and $4$ is not straightforward.  This is because middlemen at node $3$ have access to consumers  at nodes $5$ and $6$, but only one producer node $1$, on the other hand middlemen at node $4$ have access to more producers and fewer consumers.  

\begin{figure}[htbp]
\centering
\includegraphics[width=1.72in]{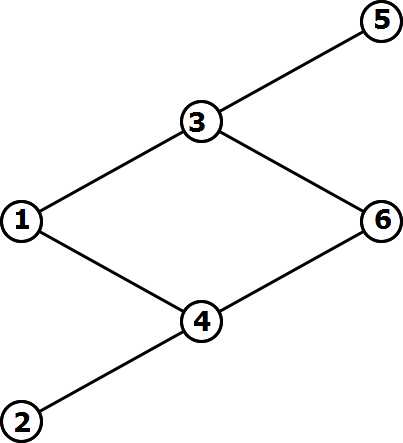}
\caption{Transaction costs at every links are  0 and the value for all buyers for obtaining a good is 1. }
\label{fig:network2}
\end{figure}

\Xomit{\bf Shorten this because we already talked about it in the intro
One could argue that because of symmetry, $3$ and $4$ should have equal payoff.  In fact, in a model like \cite{blume2009trading}, where middlemen have the all the power to make price to both producers, and consumers at the same time, one can construct an equilibrium, where the payoff of middlemen at node 3 and node 4 are the same.  In the setting that we model, middlemen cannot simultaneously  propose prices to both producers and consumers, furthermore prices are not set solely by  middlemen, they are formed through a negotiation process between producers, middlemen and consumers.  In fact we will show that in our model, middlemen at node 3 is in a better position than middlemen at node 4. The intuition is the following.  Middlemen $3$ has an competitive advantage on the consumer's side: it has access to more consumers than middlemen $4$. Thus, when holding a good, middlemen $3$ can find a consumer  easier than $4$. This had an important affect for trade in the previous round between producers and middlemen. Here even though $4$ has  access to more producers  ($1$ and $2$), but the fact that  after buying the good from these producers middleman $4$ is aware that he need to compete  and cannot get as high a surplus as  middleman 3, and thus he cannot offer good prices to the producers 1 and 2.   In other words, competition from the consumers' side has an influence back to the competition  on the producers' side, and this brings down the advantage that middleman $4$ has. 
 }

A precise prediction of the agents' behaviors depends  on the discount rate $\delta$.  The factor by which agents discount their payoff if negotiation is not successful also fundamentally influences the trade pattern. In particular, consider  trade between 1 and 4. If producers of type $1$ sell to $4$, then they might not be able to get as high a price as compared with selling to middlemen of type $3$, as they are faced with competition from $2$.  However, because the opportunity to meet with potential trade partners come randomly, depending on the discount rate, the producers might not have the incentive to wait for a trade opportunity with $3$.  Hence, given the network in Figure~\ref{fig:network0},  the pattern of trade depends on both the probability that the agents meet and their discount rate.  To make the analysis simple, and to focus on the impact of $\delta$ on the outcome of the market, we assume  the population at every node is the same, and the each pair is chosen uniformly at random. We will investigate how changing $\delta$ would influence the  trade pattern. In particular, we show that when $\delta $ is small, an equilibrium strategy is for a pair of agents who can trade (that is one has an item to sell, and the other doe not have an item),  to trade with probability 1  whenever they meet. However,  as agents are more patient ($\delta$ is close to 1), then  producers at node 1 never trade with middlemen at node 4.  This comparative analysis is summarized in Theorem~\ref{coroll0}.

Before getting to this result, for  ease of  presentation, we will introduce the following notation\footnote{Notice that here we assume all population sizes $N_i$ are the same and the distribution of selecting an edge $\pi_{ij}$ is the uniform distribution, thus $f(\delta)$ does not depend on $i,j$.}:
$$
\frac{1}{f(\delta)}:=\frac{\pi_{ij}}{2 N_i \ln(1/\delta) }.
$$
Recall the set of Bellman  equations \eqref{eq:limi1}-\eqref{eq:limzab} defining the concept of limit stationary equilibrium where the right hand side of the above equations plays an important role in the qualitative outcome of the game.  In the following we will give some comparative analysis based on $f(\delta)$. Notice that when $\delta\rightarrow 1$ , $f(\delta)$ approaches $0$, and when $\delta\rightarrow 0$ , $f(\delta)$ approaches $\infty$.

We call a stationary strategy an {\bf always trade} strategy, if for every pair of agents that are connected by a link of the network, trade occurs with probability 1, whenever they meet and one agent has an item while the other does not have one.  For a link $(i,j)$,  a stationary strategy {\bf avoids trade on} $(i,j)$ if 
when $i$, and $j$ meet, even though one agent has an item to sell and the other does not have one, they do no trade and rather wait for a future opportunity.

We summarize the comparative analysis studied in this section in the following result.
\begin{theorem}\label{coroll0}
The following hold:
\begin{itemize}
\item If $f(\delta)=1/2$, then the ``always trade'' strategy is the unique pure equilibrium,  in which case the payoff of middlemen $3$ is greater than that of middlemen $4$.
\item If $f(\delta)=1/5$, then the ``always trade'' strategy cannot be an equilibrium, and the  strategy  that avoids trade on the link $(1,4)$ and always trades on all other links is the unique pure equilibrium. At this equilibrium too, the payoff of middlemen $3$ is greater than that of  middlemen $4$

\end{itemize}

\end{theorem}

{\it Remark:} Notice that in both cases, payoff of middlemen of type $3$ is greater than that of  middlemen of type $4$.  As discussed in the introduction, this result is in contrast with models like \cite{blume2009trading}, where middlemen have all the power to set  prices to both producers and consumers {\em simultaneously}.  In our model, the fact that   middlemen need to buy an item first before selling it has important consequences. In particular, middlemen $3$ has an competitive advantage on the consumer's side: it has access to more consumers than middlemen $4$. Thus, when holding a good, middlemen $3$ can find a consumer  easier than $4$. This  in return influences trade in the previous round between producers and middlemen. Here even though $4$ has  access to more producers  ($1$ and $2$), the fact that after buying the good from these producers middleman $4$ is aware that he needs to compete and cannot get as high a surplus as  middleman 3, results in him not being able to offer as competitive a price to the producers 1 and 2.   In other words, in settings like ours, competition from the consumers' side is more important because it  has an influence back to the competition  on the producers' side.

Theorem~\ref{coroll0} illustrates another interesting phenomenon about how the discount rate $\delta$ influences the trade pattern. In particular, trade between 1 and 4 only occurs when agents are impatient enough, that is $f(\delta)$ is large.  The intuition is similar to the above argument. The advantage of $3$ over $4$ in the  consumers' market influences $4$'s ability to offer good prices to the producers. Thus, when $\delta$ is close to 1, or equivalently $f(\delta)$ is small, a producer of type 1 is better off waiting to trade exclusively with middlemen of type 3. In this case, the advantage that  $4$ has over $3$:  being connected to both producers nodes disappears, because trade between 1 and 4 will not occur.  Clearly in this case, middlemen at node $3$ are in a better position than the middlemen at node 4.

\proof{Proof of Theorem~\ref{coroll0}:}  The main idea of constructing an equilibrium or showing that a particular strategy is not an equilibrium is as follows.

First, given a strategy,  and fixing an $f(\delta)$ value, we calculate the steady state of the economy $\vec{\mu}$ that  the replicated Markov process converges to. This step can be calculated as per Lemma~\ref{theo:converge}.  For example, for an ``always trade" strategy, at node $4$, the probability of selecting a link between producers and middlemen 4 is double the probability of selecting a link  between 4 and consumers. Thus, considering the ``always-trade'' strategy the balance condition at node $4$,  we have that the rate of trading between 4 and 6 is equal to the total of trading rate between 1 and 4, and 2 and 4. This implies that the fraction of middlemen at node 4 that holds a good is twice the  fraction of middlemen at node 4 that do not hold a good. Hence, $\mu_4=2/3$. Similarly, $\mu_3=1/3$.

Second, based on the given strategy, assuming that it is an equilibrium, we obtain some constraints on the variables $z_{ij}=u_1(j)-u_0(j)-u_1(i)+u_0(i)-C_{ij}$. Specifically, in our example, $C_{ij}=0$ and trade occurs with probability 1 on every link, thus $z_{13}, z_{14}, z_{24},z_{35},z_{36}$ and $z_{46}$ are all nonnegative.
Based on the variables $z_{ij}$  and $\vec{\mu}$, we can write the expected payoff of agents $i$: $u_0(i)$ and  $u_1(i) $ according to 
(\ref{eq:limi1}-\ref{eq:limk0}). For example in the ``always-trade" strategy, we have the following:
\begin{align*}
&u_0(1)=0;  &&u_1(1)=f(\delta)(\frac{1}{3}z_{14}+\frac{2}{3}z_{14})\\
&u_0(2)=0; &&u_1(2)=f(\delta)\frac{1}{3}z_{24}\\
&u_0(3)=f(\delta)(z_{13});  &&u_1(3)=f(\delta)(z_{35}+z_{36})\\
&u_0(4)=f(\delta)(z_{14}+z_{24});  &&u_1(4)=f(\delta)(z_{46})\\
&u_0(5)=f(\delta)(\frac{1}{3}z_{35});  &&u_1(5)=1\\
&u_0(6)=f(\delta)(\frac{1}{3}z_{36}+\frac{2}{3}z_{46}); &&u_1(6)=1.\\
\end{align*}
Moreover, by definition, because  $z_{13}, z_{14}, z_{24},z_{35},z_{36}$ and $z_{46}$  are nonnegative, we have 
$$z_{ij}=u_1(j)-u_0(j)-u_1(i)+u_0(i) -C_{ij}\;\; \forall (ij) \in \mathcal{E}= \{(1,3), (1,4), (2,4),(3,5),(3,6),(4,6)\}.$$
Here, in our example, $C_ij=0$ for all links $ij$. With these we have a set of linear equations for the variables $z_{ij}$. 

The final step of the verification of an equilibrium is to solve the system of linear equations in $z_{ij}$ obtained in the previous step. Then, the given strategy is an equilibrium if and only if
\begin{itemize}
\item For a link $(i,j)$, on which we assume trade occurs with probability 1, 
$z_{ij} \ge 0$;
\item For a link $(i,j)$, on which we assume trade occurs with a probability $\lambda_{ij} \in (0,1)$, $z_{ij} = 0$; and
\item For a link $(i,j)$, on which we assume trade never occurs  $z_{ij}=u_1(j)-u_0(j)-u_1(i)+u_0(i)-C_{ij}\le 0$.
\end{itemize}

\begin{figure}[htbp]
\centering
\includegraphics[width=4.92in]{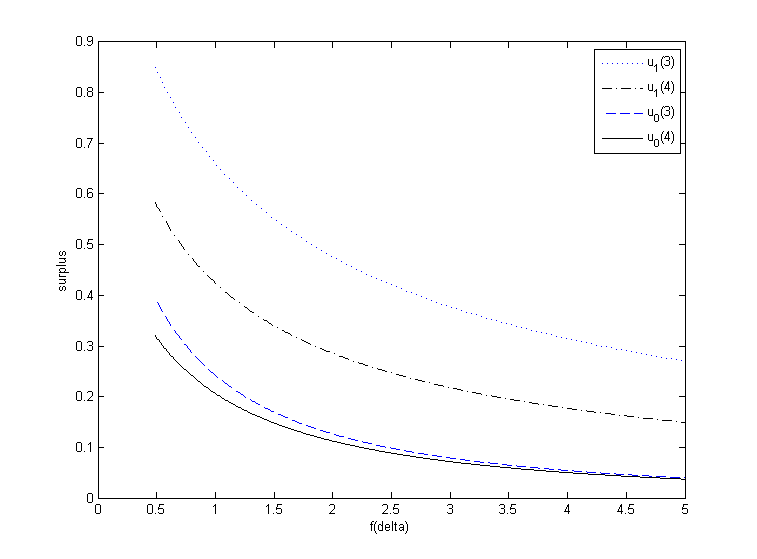}
\caption{The payoffs of the middlemen are plotted as a function of $\delta$; note that $f(\delta)$ approaches $0$ as $\delta$ approaches $1$ and infinity as $\delta$ approaches $0$.}
\label{fig:network3}
\end{figure}

Following this method, one can prove Theorem~\ref{coroll0} by numerically calculating the solution.  In particular,  with $f(\delta)=0.5$, under the ``always trade" strategy, we have the following solution, which shows that this strategy is an equilibrium:
\begin{align*} & z_{24}=0.3141; z_{14}= 0.0024; z_{13}= 0.3895; && z_{46}= 0.5782; z_{36}= 0.1911; z_{36}= 0.6537; \\
&u_0(1)=0;  u_1(1)=0.1302; &&u_0(2)=0; u_1(2)=0.0523;\\
&u_0(3)=0.1948; u_1(3)=0.4224; &&
u_0(4)=0.1582;  u_1(4)=0.2891;\\
&u_0(5)=0.1089;  u_1(5)=1; && u_0(6)=0.2246; u_1(6)=1.\\
\end{align*}
This shows that the payoff  of middlemen 3 is better than that of 4, in both states (having an item and not having an item). In fact this comparative result holds robustly, below we compute the payoff of middlemen 3 and 4 for  $f(\delta)$ ranging from $.5$ to 5, see Figure~\ref{fig:network3}. 

On the other hand, when $f(\delta)<0.49$, when using the ``always-trade" strategy we will obtain a solution in which $z_{14}$ is negative. For example, when $f(\delta) =0.2$, $z_{14}= -0.1403$, which shows that the ``always trade" strategy cannot be an equilibrium for this case. 
Recall here that $z_{14}=u_1(4)-u_0(4)-u_1(1)+u_0(1)$, which is the gain in trade when 1 sell an item to 4. If this value is negative, then 1 and 4 do not trade. 

However, with $f(\delta)=0.2$, consider the  strategy that avoids trade on the link $(1,4)$ and always trades otherwise. We will show that this is an  equilibrium.  To see this, first we observe that in this case the unique steady state of the Markov dynamic is 
$\mu_3=1/3$ and $\mu_{4}=1/2$. Similar to the previous case, we can set up a linear equation system for $z_{ij}$, where $(ij) \in \mathcal{E}= \{(1,3),(2,4),(3,5),(3,6),(4,6)\}$. Solving the system of linear equations, we get
$$ z_{24}=0.5054, z_{13}= 0.5755, z_{46}=
    0.8592, z_{36}=   0.1344, z_{35}=
    0.9399.$$
Note that these values are all positive.  Furthermore, the gain in trade between 1 and 4 is 
$$
u_1(4)-u_0(4)-u_1(1)+u_0(1)=f(\delta)\big( z_{46}-z_{24}-\frac{2}{3}z_{13}+0 \big)=-0.0299\cdot f(\delta)<0.
$$
This shows that, indeed, 1 and 4 do not have an incentive to trade. It is also clear here that, in this case, middlemen of type 3 have a higher payoff than middlemen of type 4.

Lastly, by considering all other pure strategies, we can conclude that for $f(\delta)=0.5$ the ``always trade" strategy, and for $f(\delta)=0.2$ ``avoiding trade on the link $(1,4)$ and always trades on all other links" are the unique pure equilibrium, respectively. 


\Xomit{
\begin{figure}[htbp]
\centering
\includegraphics[width=4in]{examples.png}
\caption{Network}
\label{fig:network2}
\end{figure}

\subsection{Star networks}
The first one is a star network where a single node of middlemen is connected to $n_1$ nodes of  producers and $n_2$ nodes of consumers. See figure~\ref{fig:network2}. 

At stationary equilibrium: $\alpha=n_1/(n_1+n_2)$ fraction of the middlemen hold the good; $1-\alpha=n_2/(n_1+n_2)$ are empty handed and wait to buy. 

Let $x$ be the expected payoff of producer.  Because of  equation?? then the ratio of  the expected payoff of middlemen when not holding to $x$
is the the same as the ratio of their probability being selected as a proposer. 	

a good is : 

\subsection{Trees}
}

\Xomit{
In particular, when $\delta$ approach 1, then $f(\delta)$ approaches 0, and as seen in Section???? in this case agents are patients and willing to wait for the shortest trade route. (Mathematically, as $f(\delta)$ approaches 0, we show that $\max \{z_{ij},0\}$ )  approaches $0$ for all $i,j$.)On the other hand,  when $\delta$ approaches 0, agent discount their future payoff aggressively, this means that agents are very impatient.  It is clear then that from $f(\delta)$ approaches infinity, and the expected payoff of all agents approaches 0. This is intuitive, because in our model as the economy gets large, in every period the probability that an agents is selected in the trade is getting small, thus as $\delta$ approaches 0, agents  discounts almost all their payoff because of waiting. 
}

\section{Comparative Studies With Patient Agents}\label{comparative}
In this section we will focus our comparative studies on the case when the discount factor $\delta$ approaches 1, which we sometimes refer to as when ``agents are being patient'' or  the case of ``vanishing bargaining friction." In many cases, by considering this limiting case, we are either able to give a closed form characterization of the equilibrium, or present robust and general  properties of any equilibrium.


We start by showing that  as agents become patient, they will choose the cheapest trading routes (i.e. the routes with the smallest transaction costs) and intermediary fees per transaction will approach 0.  We then give a closed form characterization  for the  equilibrium  in a simple network containing two links. This shows how endogenous delay emerges even in this simple network. Finally, we use these two results to give an example of how changing the transaction costs in a network can have a fundamental and counter intuitive impact on agents' payoffs.

\subsection{Preference for Cheapest Trade Routes} \label{sec:efficiency}

\begin{theorem}\label{corol1}
Given a producer $p$, a consumer $c$, and any $\epsilon>0$,  there exists $\delta^*$, such that for all $\delta>\delta^*$ and at any equilibrium the following is true.  If  $\lambda_{pm}>0$ and $\lambda_{mc}>0$ for a middlemen $m$, that is trade occurs along the route $p\rightarrow m \rightarrow c$, then  the cost $C_{pm}+C_{mc}$ is the smallest among all trading routes between $p$ and $c$.  Furthermore, the total surplus of producer $p$ and consumer $c$ satisfies $u_1(p)+u_0(c)\ge V_v-(C_{pm}+C_{mc})-\epsilon$. 
\end{theorem}

\proof{Proof of Theorem \ref{corol1}:} See Appendix~\ref{app:corol1}. 

{\it Remark:} The above result demonstrates a global-level efficiency that emerges in the equilibria of the local non-cooperative bargaining scheme if agents are patient enough: edges that are not along a cheapest path from any producer and consumer pair are never used, and middlemen who have no edges along a cheapest path from any producer and consumer pair see no trade.

 Furthermore, $u_1(p)+u_0(c)\ge V_v-(C_{pm}+C_{mc})-\epsilon$ implies that the total surplus of producer $p$ and consumer $c$ is almost the entire trade surplus on the path $p \rightarrow m \rightarrow c$.  This implies that as $\delta$ approaches 1, the intermediary's fee that $j$ charges approaches 0.  We discussed the intuition for this property in the introduction (Section~\ref{intro}). One important feature of our model that leads to this result is the fact that middlemen are long-lived and do not consume the good, while producers and consumer have limited supply and demand. Middlemen are eager to buy and sell quickly, while producers and consumers can wait because the discount rate $\delta$ is close to 1.  It can be seen mathematically in the proof that when  $\delta$ approaches 1, $z_{pm}$ and $z_{pc}$ approach 0. The term $z_{pm} + z_{mc}$ captures the gain of surplus for middlemen $m$ after buying an item from $p$ and selling it to $c$, assuming in both transactions, $m$ is the proposer.


\begin{figure}[htbp]
\centering
\includegraphics[width=1.52in]{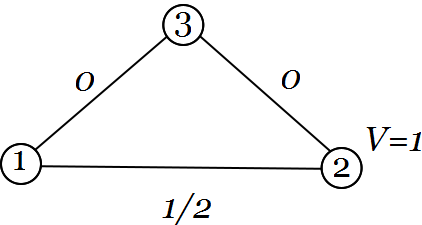}
\caption{The trading network used to explore the choice of trade routes for $\delta \in (0,1)$.}
\label{fig:triangle}
\end{figure}

Also notice that the selection of the cheapest routes does not hold when agents discount their payoff by $\delta$ much smaller than 1. To see this consider the network illustrated in Figure~\ref{fig:triangle}, 
where there are two paths connecting producer 1 and consumer 2. 
The direct link has a cost of 1/2, and both links connecting to middleman 3 costs 0.
Hence, from Theorem~\ref{corol1}, when $\delta$ is large enough trade will only occur through the middleman. However, when $\delta$ is small enough, 
even though the surplus of the direct trade is $1-1/2 =1/2$, trade will happen on link $(1, 2)$ when it is selected instead of waiting  to trade through the middlemen of type $3$.  In particular, assume  the population at each node is the same, say $N$, and each link is selected uniformly at random so that $\pi_{ij}=1/3$ for all $(i,j)$.

\begin{figure}[htbp]
\centering
\includegraphics[width=5in]{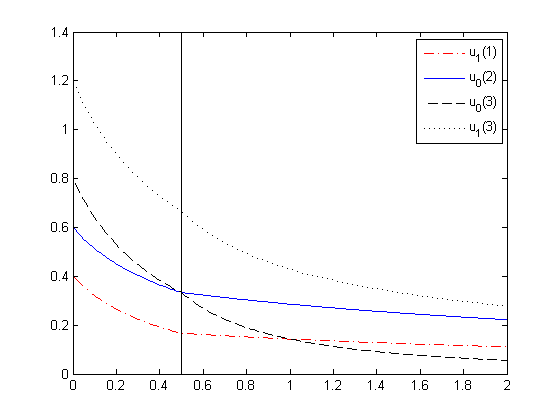}
\caption{Agent payoffs as a function of $f(\delta)$.}
\label{fig:triangle2}
\end{figure}

Similar to the computation used in Theorem~\ref{coroll0}, let 
$$
\frac{1}{f(\delta)}:=\frac{\pi_{ij}}{2 N \ln(1/\delta)}= \frac{1}{6 N \ln(1/\delta) }.
$$
It is then straightforward  to check that the following is a limit stationary equilibrium.
\begin{itemize}
\item ``Always trade" if  $f(\delta)\geq 1/2$; 
\item Always trade on $(1,3)$, $(3,2)$ and avoid trade on $(1,2)$, if  $f(\delta) < 1/2$, which corresponds to $\delta \geq \exp\left(-\tfrac{1}{3N} \right)$.
\end{itemize}
In Figure~\ref{fig:triangle2},  we plot the payoff of 1, 2, and 3 at this equilibrium as $f(\delta)$ varies from 0 to 2; these are decreasing as  $f(\delta)$ increases. 

\subsection{Endogenous Delay}\label{sec:delay}
\begin{figure}[htbp]
\centering
\includegraphics[width=2.5in]{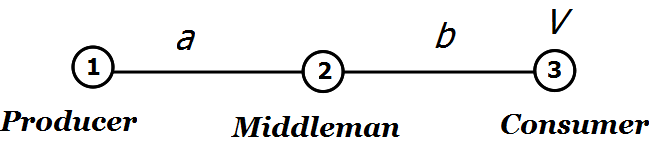}
\caption{A simple network to illustrate endogenous delay.}
\label{fig:2a}
\end{figure}

We now consider a simple network that consists of two links illustrated in Figure~\ref{fig:2a}, 
which was again discussed in the introduction (Section~\ref{intro}).  This network represents the simplest example where producers and consumers cannot trade directly. We fully characterize the limit stationary equilibrium in this example, which will be shown to be  unique. Even in this simple network, we observe an interesting phenomenon of endogenous delay as part of the equilibrium.  This is counterintuitive since in a full information dynamic bargaining model like ours, delay in trade does not enable agents to learn any new information, but  only decreases the total surplus of trade. Therefore, the network structure and the combination of incentives of long-lived and short-lived agents  are the main sources causing this inefficiency in bargaining.

Assume $a$ and $b$ are transaction costs of the first and second link, also let  $V$ be the value of the consumption of the good; without loss of generality, we will insist that trade is favorable so that $V> a+b$.  The probabilities of using the links are then $\pi_{12}$ and $\pi_{23}$.   We assume the population sizes at every node is equal, and without loss of generality, we assume\footnote{Following the proof of this result, it will become clear that this assumption does not result in any loss of generality when agents are patient.}  $N_1=N_2=N_3=1$. We will show that in this simple network, the stationary equilibrium is unique, and we characterize the condition under which agents do not trade immediately. 
\begin{theorem}\label{theo:twohops}
In the limit of $\delta \rightarrow 1$, that is, when the agents are patient, there is always a unique limit stationary equilibrium. Furthermore, 
if $V\ge \left(1+\tfrac{\pi_{12}}{\pi_{12}+\pi_{23}}\right) a+ b=:\bar{V}$, then trade always happens, otherwise there is a delay. The probability of trade on link $(1,2)$, $\lambda_{12}$, the probability of trade  on link $(2,3)$, $\lambda_{23}$, and the equilibrium state and the payoffs of the middleman are given by
\begin{align*}
\lambda_{23} & = 1,  & \lambda_{12} & = 
\begin{cases}
1 & \text{if } V\ge \bar{V} \\
\frac{\pi_{23} (V-b-a)}{\pi_{12} (2a+b-V)} & \text{otherwise}
\end{cases}, \\
\mu_2 &= 
\begin{cases}
\frac{\pi_{12}}{\pi_{12}+\pi_{23}} & \text{if } V\ge  \bar{V} \\
\frac{V-b-a}{a} & \text{otherwise}
\end{cases}, & \\
u_1(2) & = 
\begin{cases}
\frac{(V-b)\left(2-\frac{\pi_{12}}{\pi_{12}+\pi_{23}}\right)-a}{1+\frac{\pi_{12}\pi_{23}}{(\pi_{12}+\pi_{23})^2}} & \text{if } V\ge  \bar{V} \\
a & \text{otherwise}
\end{cases}, &
u_0(2) & =
\begin{cases}
\frac{V- \left(1+\frac{\pi_{12}}{\pi_{12}+\pi_{23}}\right) a- b }{1+\frac{\pi_{12}\pi_{23}}{(\pi_{12}+\pi_{23})^2}} &  \text{if } V\ge  \bar{V} \\
0 & \text{otherwise}
\end{cases}, \\
u_0(3) & = 
\begin{cases}
\frac{(V-b)\left(2-\frac{\pi_{12}}{\pi_{12}+\pi_{23}}\right)-a}{\frac{\pi_{12}+\pi_{23}}{\pi_{12}}+\frac{\pi_{23}}{\pi_{12}+\pi_{23}}} & \text{if } V\ge  \bar{V} \\
1-b-a & \text{otherwise}
\end{cases}, &
u_1(1) & = 
\begin{cases}
\frac{V- \left(1+\frac{\pi_{12}}{\pi_{12}+\pi_{23}}\right) a- b }{\frac{\pi_{12}+\pi_{23}}{\pi_{12}}+\frac{\pi_{23}}{\pi_{12}+\pi_{23}}} &  \text{if } V\ge  \bar{V} \\
0 & \text{otherwise}.
\end{cases}
\end{align*}
\end{theorem}
{\it Remark:}  As discussed in the introduction (Section~\ref{intro}), this can be compared with a model where each node consists of a single agent, where the sunk cost problem causes market failure, and no trade is the unique equilibrium.  Here, however, we show that when the stock at the middlemen node is small, the search friction for a consumer to find a middlemen that owns an item increases the middlemen's bargaining power, and this reestablishes trade with a positive probability.   

From Theorem~\ref{theo:twohops}, we also see that trade always occurs on link $(2,3)$ but can be delayed at link $(1,2)$. Since the consumer is at the other end of link $(2,3)$, it stands to reason that there is no delay in the trade. However, at link $(1,2)$, any sale of the item results in a decreased likelihood of the trade at the same link (in the near future) owing to the search mechanism, and this opportunity cost introduces the delay in trade. Note also that with a delay in trade, the producer obtains no surplus! We will revisit this effect in the next section when we discuss the impact of the delay on the share of the surplus between the agents. Trade gets delayed when the value of the good is below a specific threshold. From the proof one can discern that the additional penalty term in the threshold is the product of the transaction cost at link $(1,2)$ and the stationary probability that the middleman possesses the good.

\proof{Proof of Theorem~\ref{theo:twohops}:} See Appendix~\ref{app:twohops}. 


%
%

\subsection{Share Of Surplus}\label{sec:surplusdiv}
Lastly, we consider the imbalance between the surplus of producers and consumers as a result of our decentralized trade model.  

\begin{figure}[htbp]
\centering
\includegraphics[width=2.5in]{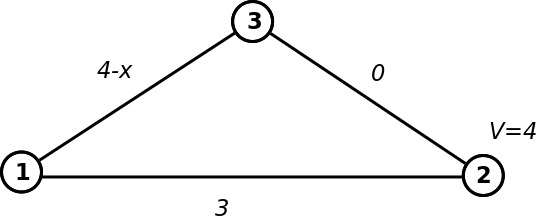}
\caption{Triangle network with transaction costs.}
\label{fig:3}
\end{figure}

Consider the following (also simple) network, where node 1 represents producers, node 2 represent consumers and node 3 represents middlemen, as illustrated in Figure~\ref{fig:3}.  Again without loss of generality, we assume  that$N_1=N_2=N_3=1$.
 We also assume that in our bargaining model, every link is selected uniformly at random, that is $\pi_{ij}=1/3$ for all $i\neq j$. Assume the consumer's valuation for the good is  $V_2=4$, and the transaction costs are the following:
 $C_{12}=3$, $C_{32}=0$ and $C_{13}=4-x$. We will investigate the equilibrium as $x$ changes. 
As $x$ increases, the transaction cost between $1$ and $3$ decreases, making the total trade surplus $\max\{4-3,4-(4-x)\}=\max\{ 1,x\}$ increase.


The surplus of producers in this example is understood as the payoff of agents at node 1 when owning an item: $u_1(1)$.
On the other hand,  the surplus of consumers in this example is the payoff of agents at node 2 when not owning an item: $u_0(2)$. According to the analysis in Section~\ref{sec:efficiency}, as the discount rate $\delta$ approaches 1, trade will only goes through the cheapest route. Let $\bar{C}$ be the cost of this route, as seen in Section~\ref{sec:efficiency} we also  have
$$
\lim_{\delta\rightarrow 1} u_1(2)-u_0(2) -(u_1(1)-u_0(1)) = \bar{C}. 
$$
This is equivalent to 
$$
\lim_{\delta\rightarrow 1} u_0(2) + u_1(1) = V_2-\bar{C}. 
$$

As shown in Section~\ref{sec:efficiency} this means that  the total surplus of a producer and a consumer approaches the total trading surplus, and  for every transaction, middlemen only make a vanishing amount of fee.      As discussed in the introduction, this is because in our model, we assume  producers and consumers are short-lived, while middlemen are long-lived and has to earn money by flipping the item. Thus, as $\delta$ approaches 1, while producers and consumers are patient, middlemen are eager to buy and sell quickly.

Now, in the example above,  when considering the equilibrium payoff as $\delta$ approaches 1, we have  if $x<1$ that producers and consumers will trade directly, and in this case producers and consumers equally share the surplus, giving them a surplus of $\tfrac{4-3}{2}=\tfrac{1}{2}$.  

\begin{figure}[htbp]
\centering
\includegraphics[width=4in]{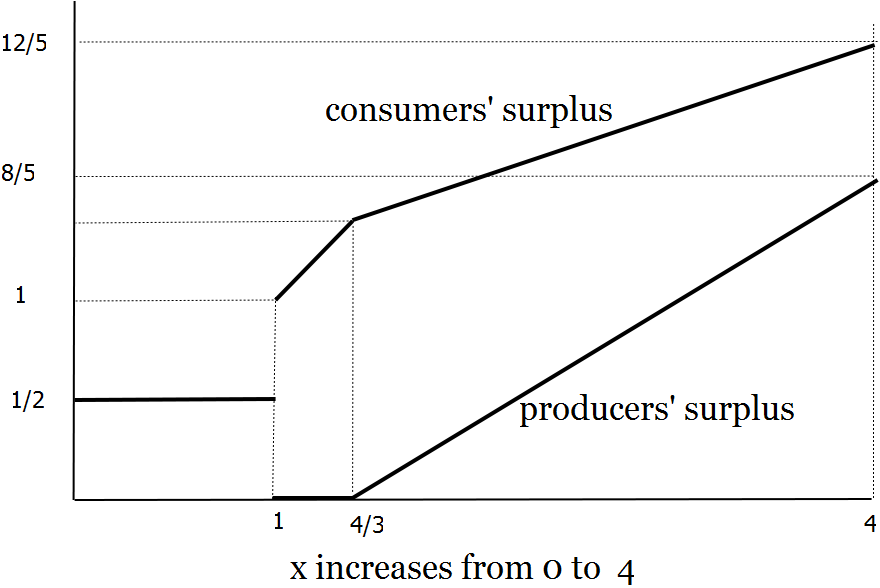}
\caption{Surplus of producer and consumers as $x$ increases from 0 to 4}
\label{fig:4}
\end{figure}

On the other hand, if $x>1$, then direct trade between producers and consumers is too expensive, and trade will go through middlemen at  node 3. In the latter case, we will use the analysis in 
Section~\ref{sec:delay}  to compute the equilibrium payoff, and we have
\begin{enumerate}
\item $1<x<4/3$: {Seller's surplus,} $u_1(1)=0$ and
consumer's surplus $u_0(2)=x$, so that the consumers get all the trade surplus;
\item  $4/3\le x\le 4$: {Seller's surplus,} $u_1(1)=\frac{4-3/2(4-x)}{5/2}=\frac{3x-4}{5}$ and 
consumer's surplus $u_0(2)=\frac{2x+4}{5}$.
\end{enumerate}
This is illustrated in Figure~\ref{fig:4}.

Even in  this simple network, we observe quite an interesting phenomenon: there is a discontinuous shift in the trading pattern occurring in the network.  If $x$ increases from $1$ to $4/3$, so that the transaction cost between 1 and 3, $C_{13}=4-x$, decreases,  then the total surplus between producers and consumers  increases, but producers are actually worse off because  of this shift in the  market structure. This also highlights how local adjustments by the producers could leave them in a worse-off position. Note that when $x=1$, at least two limit stationary exist: trade directly or via middlemen.

This example captures an interesting and  counterintuitive phenomenon: as the transaction cost towards middlemen decreases, producers  can be worse off, because the high cost of direct trading makes consumers refuse to trade directly and prefer to trade through middlemen. For example, in many supply chain networks, as these global networks get large, producers and consumers do not trade directly and several types of organizations emerge as  middlemen. In many cases such as in coffee industry, producers (coffee farmers) obtain a very small fraction of surplus because there are too many middlemen in the supply chain network. See for example \cite{Bacon} for a related empirical analysis  of  the coffee global supply chain and  the recent shift in its market structure.

\section{Conclusions And Future Work}\label{conc}
In this paper we considered non-cooperative local bargaining over a trading network with a single type of good. In the limiting scenario of many agents, we showed the existence of a limit stationary equilibrium that can be characterized by a combination of the stationary probability of a trade happening on each link, the stationary distribution of the agents' possessing the good, and the stationary payoffs of the agents. We then showed that when agents are patient enough, this limiting equilibrium can exhibit global efficiency.  We applied this concept to several simple network structures to study the impact of the network on the bargaining power and surplus of all agents. In future work we plan to extend the results to more general networks, to include the analysis of losing or damaging the good and to rigorously connect the equilibria in the finite-player game to the limit stationary equilibria.


\section{APPENDIX}


\subsection{Proof of Lemma~\ref{theo:converge}}\label{sec:proofMarkov}

Since the transition matrix of the Markov process $\{\tilde{X}_v^k(t): v \in \mathcal {V}\}$ satisfies a Lipschitz condition, by an application of Kurtz's Theorem \cite[Th. 2.1, Chapter 11]{EthierKurtz2005}, we obtain a differential equation for the limiting process. The globally asymptotically stable state of the differential equation is a continuous function from the non-negative reals to $\prod_{p\in\mathcal{P}}[0,N_p]\times\prod_{m\in\mathcal{M}}[0,N_m]\times\prod_{c\in\mathcal{C}}[0,N_c]$. The limiting processes\footnote{Even though Kurtz's Theorem applies only for finite time-horizons, the compact setting of the scaled processes and the well-behaved nature of the differential equation above allow us to analyze the convergence of the stationary solutions as well.} are given as follows for all $t\geq 0$,
\begin{align}
& \forall p\in\mathcal{P},\quad x_p(t)  \equiv N_p; \qquad\qquad
\forall c\in\mathcal{C},\quad  x_c(t)  \equiv 0; \notag \\
& \forall m \in \mathcal{M}, \quad \frac{d x_m(t)}{dt}  = \left(1-\frac{x_m(t)}{N_m}\right)\sum_{p\in \mathcal{P}: (p,m) \in \mathcal{E}_2} \pi_{pm} \lambda_{pm}  - \frac{x_m(t)}{N_m} \sum_{c\in \mathcal{C}: (m,c) \in \mathcal{E}_3} \pi_{mc}\lambda_{mc}  \label{eq:balance2} \\
& \qquad \qquad = \sum_{p\in \mathcal{P}: (p,m) \in \mathcal{E}_2} \pi_{pm} \lambda_{pm}  - \frac{x_m(t)}{N_m} \left(\sum_{p\in \mathcal{P}: (p.m) \in \mathcal{E}_2} \pi_{pm} \lambda_{pm} +  \sum_{c\in \mathcal{C}: (m,c) \in \mathcal{E}_3} \pi_{mc}\lambda_{mc} \right). \label{eq:diffeq}
\end{align}
Using a quadratic Lyapunov function (square of the distance to the equilibrium point) it follows\footnote{The details are omitted as this is a standard technique for the linear dynamics in \eqref{eq:diffeq}.} that there is a unique and globally asymptotically stable equilibrium point that is given by
\begin{align*}
\forall p\in\mathcal{P},\quad x_p^* & = N_p;\qquad\quad
\forall c\in\mathcal{C},\quad x_c^* = 0; \\
\forall m \in \mathcal{M}, \quad x_m^* & = N_m \frac{\sum_{p\in \mathcal{P}: (p,m) \in \mathcal{E}_2} \pi_{pm} \lambda_{pm} }{\sum_{p\in \mathcal{P}: (p,m) \in \mathcal{E}_2} \pi_{pm} \lambda_{pm} +  \sum_{c\in \mathcal{C}: (m,c) \in \mathcal{E}_3}\pi_{mc} \lambda_{mc}}.
\end{align*}
Therefore, the fraction of agents with the good satisfies
\begin{align*}
\forall p\in\mathcal{P},\quad \mu_p & = 1;\qquad\quad
\forall c\in\mathcal{C},\quad \mu_c = 0; \\
\forall m \in \mathcal{M}, \quad \mu_m & = \frac{\sum_{p\in \mathcal{P}: (p,m) \in \mathcal{E}_2} \pi_{pm} \lambda_{pm} }{\sum_{p\in \mathcal{P}: (p,m) \in \mathcal{E}_2} \pi_{pm} \lambda_{pm} +  \sum_{c\in \mathcal{C}: (m,c) \in \mathcal{E}_3}\pi_{mc} \lambda_{mc}}.
\end{align*}
Note that setting the right-hand side of \eqref{eq:balance2} to zero, yields the balance condition \eqref{eq:balance}.

For each $k$, it is easy to see that the Markov process is irreducible and has finite states, and so is positive recurrent. Thus, owing to the compact setting, the stationary measures of the scaled state processes converge to the point mass on the equilibrium point as $k$ increases without bound, see \cite{Benaim:2011fk}. 

\subsection{Proof of  Theorem~\ref{theo:existence}}\label{sec:proofexist}

We need to show that there exists $(\vec{\lambda}, \vec{\mu}, \vec{u}, \vec{z})$ satisfying the following conditions:
\begin{enumerate}
\item Convergence: given the trading dynamics defined by $\vec{\lambda}$, the replicated economy converges to the steady state $\vec{\mu}$. According to Lemma~\ref{theo:converge}, the convergence occurs and the dependence of $\vec{\mu}$ on $\vec{\lambda}$ is given by \eqref{eq:dynamic1}-\eqref{eq:dynamic2};
\item Payoff-state consistency: $\vec{u}$ and $\vec{z}$ need to satisfy equations \eqref{eq:limi1}-\eqref{eq:limzab}; and 
\item Payoff-dynamic  consistency: if $z_{ij}>0$ then  $\lambda_{ij}=1$; if $z_{ij}< 0$ then   $\lambda_{ij}=0$; and if $z_{ij}=0$, then $0\leq \lambda_{ij}\leq1$.
\end{enumerate}

Given $(\vec{\lambda}, \vec{\mu}, \vec{u})$, using equations \eqref{eq:limi1}-\eqref{eq:limzab} and \eqref{eq:dynamic1}-\eqref{eq:dynamic2} as well as the payoff-dynamic consistency above, we can obtain the following correspondence $(\vec{\Lambda}, \vec{\mu}^\prime, \vec{u}^\prime)$, which we can write as follows,
$$ F(\vec{\lambda}, \vec{\mu},  \vec{u})= (\vec{\Lambda}, \vec{\mu}^\prime, \vec{u}^\prime),$$
where
\begin{align*}
\forall p\in\mathcal{P}\quad \mu^\prime_p & = 1, \\
\forall c\in\mathcal{C}\quad \mu^\prime_c & = 0, \\
\forall m \in \mathcal{M} \quad \mu^\prime_m & = \frac{\sum_{p\in \mathcal{P}: (p,m) \in \mathcal{E}_2} \pi_{pm} \lambda_{pm} }{\sum_{p\in \mathcal{P}: (p,m) \in \mathcal{E}_2} \pi_{pm} \lambda_{pm} +  \sum_{c\in \mathcal{C}: (m,c) \in \mathcal{E}_3}\pi_{mc} \lambda_{mc}},
\end{align*}
and for all $(i,j)\in \mathcal{E}$
\begin{align*}
\Lambda_{ij}=&\{1\} \text{ if } u_1(j)-u_0(j) -(u_1(i)-u_0(i))-C_{ij}>0,\\
\Lambda_{ij}=&\{0\} \text{ if } u_1(j)-u_0(j) -(u_1(i)-u_0(i))-C_{ij}<0,\\
\Lambda_{ij}=&[0,1] \text{ if  }u_1(j)-u_0(j) -(u_1(i)-u_0(i))-C_{ij}=0.
\end{align*}
Furthermore,
\begin{align*}
\begin{split}
\forall p \in\mathcal{P}  \quad u^\prime_0(p) & = 0,\\
\quad u^\prime_1(p)  &= \sum_{c: (p,c) \in \mathcal{E}_1}  \frac{\pi_{pc}}{2 N_p \ln(1/\delta) } \max\{z_{pc},0 \} 
+ \sum_{m: (p,m) \in \mathcal{E}_2}  \frac{\pi_{pm}}{2 N_p \ln(1/\delta) } (1-\mu_m) \max\{z_{pm}, 0 \} ,
\end{split}\\
\forall m \in \mathcal{M} \quad u^\prime_0(m) & = \sum_{p: (p,m) \in \mathcal{E}_2} \frac{\pi_{pm}}{2 N_m \ln(1/\delta)} \max\{z_{pm},0\}, \\
\forall m \in \mathcal{M} \quad u^\prime_1(m) & = \sum _{c: (m,c) \in \mathcal{E}_3} \frac{\pi_{mc}}{2 N_m \ln(1/\delta) } \max\{z_{mc},0 \}, \\
\begin{split}
\forall c \in \mathcal{C} \quad u^\prime_0(c) & = \sum_{p: (p,c) \in \mathcal{E}_1}  \frac{\pi_{pc}}{2 N_c \ln(1/\delta) }\max\{z_{pc},0 \} 
+ \sum _{m: (m,c) \in \mathcal{E}_3}  \frac{\pi_{mc}}{2 N_c \ln(1/\delta) } \mu_m \max\{z_{mc},0 \},\\
\quad u'_1(c) & = V_c, \text{ where }
\end{split}  \\
 z_{ij} &= \Big(u_1(j) - u_0(j) - \Big(u_1(i)-u_0(i) \big)\Big)-C_{ij} \;\;\; \forall (i,j) \in \mathcal{E}_1 \cup \mathcal{E}_2 \cup \mathcal{E}_3.
 \end{align*}


It is straightforward to check that the function $F(\cdot,\cdot,\cdot)$ above satisfies all the requirements for Kakutani's fixed-point theorem. The domain is a non-empty, compact and convex subset of a finite-dimensional Euclidean space. The mapping/correspondence has a closed-graph: since the mappings from $\vec{\lambda}$ to $\vec{\mu}^\prime$ and $\vec{\mu}$ to $\vec{u}^\prime$ are single-valued and continuous, we only need to satisfy this for the mapping from $\vec{u}$ to $\vec{\Lambda}$. For any sequence $(\vec{u}_n,\vec{\Lambda}_n)$ (in the domain) such $\lim_{n\rightarrow\infty} (\vec{u}_n, \vec{\Lambda}_n)=(\vec{u},\vec{\Lambda})$, it is easy to see that $\vec{\Lambda}$ must lie in the image of $\vec{u}$. Finally, the image of any point in the domain is non-empty, closed and convex. Therefore, there must be  a fixed-point, and furthermore, by definition, any fixed point of this mapping is a limit stationary equilibrium.

\subsection{Proof of Theorem~\ref{corol1}}\label{app:corol1}
Consider equations \eqref{eq:limi1}, \eqref{eq:limj0}, \eqref{eq:limj1} and \eqref{eq:limk0}. As $\delta$ approaches $1$, the $\log(1/\delta)$ term approaches $0$. Since $u_1(p) \in [0, \max_{c\in\mathcal{C}} V_c]$ for all $i\in \mathcal{P}$, and $u_0(c) \in [0, \max_{c\in\mathcal{C}} V_c]$ for all $c\in \mathcal{C}$, it has to be that given any $\epsilon>0$, there exists $\delta^*$ such that for all $\delta>\delta^*$, we have 
\begin{align*}
z_{pc} & \leq \epsilon \quad \forall (p,c) \in \mathcal{E}_1,\\
z_{pm} & \leq \epsilon \quad \forall (p,m)\in\mathcal{E}_2, \\
z_{mc} & \leq \epsilon \quad \forall (m,c)\in\mathcal{E}_3.
\end{align*}
Now consider a pair of agents, one producer $p$ and consumer $c$. We have three cases then:
\begin{enumerate}
\item All trade routes from $p$ to $c$ have to visit some middleman. Let $m\in\mathcal{M}$ be one such middleman so that $(p,m) \in \mathcal{E}_2$ and $(m,c) \in \mathcal{E}_3$. The inequalities above then imply the following:
\begin{align*}
u_1(m) - u_0(m) & \geq u_1(c) - u_0(c) - C_{mc} -\epsilon,\\
u_1(m) - u_0(m) & \leq u_1(p) - u_0(p) + C_{pm} +\epsilon.
\end{align*}
These with $u_0(p)=0$ and $u_1(c)=V_c$ imply
\begin{align*}
u_1(p)+u_0(c) \geq V_c - C_{pm} - C_{mc}-2\epsilon.
\end{align*}
Note that this inequality holds for every $m\in \mathcal{M}$ that lies along a trade route from $p$ to $c$.
Therefore,
\begin{align*}
u_1(p) + u_0(c) \geq V_c - \min_{\{m: (p,m) \in \mathcal{E}_2 \text{ and } (m,c) \in \mathcal{E}_3\}} \Big( C_{pm} + C_{mc}\Big) -2\epsilon.
\end{align*}
Since $\epsilon$ can be chosen arbitrarily small, thus for any middleman $m$ who is not on a smallest transaction cost path from $p$ to $c$, we can choose $\delta$ close enough to 1 such that  either $z_{pm}$ or $z_{mc}$ is strictly negative and so no trade can occur on the corresponding edge;

\item Notice that the same argument also works for the case, where if in addition to the middlemen, there also exists a direct link between $p$ and $c$. Then
\begin{align*}
u_1(p) + u_0(c) \geq V_c - \min\bigg(C_{pc}, \min_{\{m: (p,m) \in \mathcal{E}_2 \text{ and } (m,c) \in \mathcal{E}_3\}} \Big( C_{pm} + C_{mc}\Big)\bigg).
\end{align*}
Again it is clear that no trade occurs over links that are not part of a smallest transaction cost path from $p$ to $c$;
\item If $p$ and $c$ only have a direct route between them, then that is the only route via which trade can occur between this producer and consumer pair. Also, if no routes exist between $p$ and $c$, then obviously no trade occurs between these two agents.
\end{enumerate}

\subsection{Proof of Theorem~\ref{theo:twohops}}\label{app:twohops}

The equilibrium equations for this case are as follows:
\begin{align*}
u_1(1) & = \frac{\pi_{12}}{2\ln(1/\delta)} (1-\mu_2)\max\{z_{12},0\}, &
u_0(2) & = \frac{\pi_{12}}{2\ln(1/\delta)} \max\{z_{12},0\},\\
u_1(2) & = \frac{\pi_{23}}{2\ln(1/\delta)}\max\{z_{23},0\}, &
u_0(3) & = \frac{\pi_{23}}{2\ln(1/\delta)} \mu_2 \max\{z_{23},0\},\\
z_{12} & = (u_1(2)-u_0(2)-u_1(1))-a,  & 
z_{23} & = (V-u_0(3) - u_1(2) + u_0(2))-b,\\
\lambda_{12} & \in 
\begin{cases}
\{1\} & z_{12} > 0\\
\{0\} & z_{12} < 0\\
[0,1] & z_{12}=0
\end{cases}, \quad 
\lambda_{23}  \in 
\begin{cases}
\{1\} & z_{23} > 0\\
\{0\} & z_{23} < 0\\
[0,1] & z_{23}=0
\end{cases},  & \mu_2 & = \frac{\pi_{12} \lambda_{12}}{\pi_{12} \lambda_{12} + \pi_{23} \lambda_{23}}.\\
\end{align*}
From the above is clear that $u_1(1)=(1-\mu_2) u_0(2)$ and $u_0(3) = \mu_2 u_1(2)$. Substituting these we get
\begin{align*}
z_{12}  = \big(u_1(2) - (2-\mu_2) u_0(2) \big) -a, \qquad
z_{23}  = \big( V - (1+\mu_2) u_1(2) + u_0(2) \big) - b.
\end{align*}

First consider the assumption that trade occurs with probability one on both links, i.e., $\lambda_{12}=\lambda_{23}=1$. This then implies that $\mu_2=\tfrac{\pi_{12}}{\pi_{12}+\pi_{23}}$, $z_{12}, z_{23} \geq 0$ and we can substitute them directly into the equations for the payoffs. We then obtain the following linear equations in $u_0(2)$ and $u_1(2)$,
\begin{align*}
u_0(2)  = \frac{u_1(2) -a}{\frac{2\ln(1/\delta)}{\pi_{12}}+ (2-\mu_2)}, \quad
u_1(2)  = \frac{u_0(2) + V -b}{\frac{2\ln(1/\delta)}{\pi_{23}}+(1+\mu_2)}.
\end{align*}
We can take limits in the equations above as $\delta$ goes to $1$ (along an appropriate subsequence) to get
\begin{align*}
u_0(2)(2-\mu_2)  = u_1(2) -a, \qquad
u_1(2)  (1+\mu_2)  = u_0(2) + V -b.
\end{align*}
Since the $\ln(1/\delta)$ terms vanish in the limit of $\delta$ going to $1$, so would the impact of the different population sizes. The unique solution of the above system of linear equations is 
\begin{align}
u_1(2)  = \frac{(2-\mu_2) (V-b) -a}{1+\mu_2 - \mu_2^2}, \quad
u_0(2)  = \frac{V-(1+\mu_2)a-b}{1+\mu_2 - \mu_2^2}.
\label{eq:zazb}
\end{align}
It is easily seen that $u_1(2)\geq 0$ and $u_0(2)\geq 0$ if and only if $V \geq (1+\mu_2) a - b=\bar{V}$, and at the equilibrium $z_{12}=z_{23}=0$. 

For the remainder assume that $V < \bar{V}$. Consider the case that $\lambda_{23}=1$ and $0< \lambda_{12} < 1$. This then implies that $z_{12}=0$, $z_{23}\geq 0$ and $\mu_2 = \tfrac{\pi_{12} \lambda_{12}}{\pi_{12} \lambda_{12}+ \pi_{23}}$. Again taking a limit of $\delta$ going to $1$ (along an appropriate subsequence), we also get $z_{23}=0$. If we solve \eqref{eq:zazb}, then the calculated $u_0(2)$ will be negative which then implies that at the equilibrium $u_0(2)=0$; note that $z_{12}=0$ for $\delta <1$ also yields the same conclusion. Now it follows that
\begin{align*}
u_1(2)=a, \; \mu_2=\frac{V-b-a}{a}, \text{ and } \lambda_{12}=\frac{\pi_{23} (V-b-a)}{\pi_{12} (2a+b-V)} \in (0,1).
\end{align*}
Since $V < \bar{V}=\left(1+\tfrac{\pi_{12}}{\pi_{12}+\pi_{23}}\right) a + b$, it also follows that $V < 2 a + b$ which ensures $\mu_2 \leq 1$. Similarly, one can verify that $\lambda_{12} \in (0,1)$. Since the consistency conditions are met, we have an equilibrium. 

The uniqueness of the solution in both cases also proves that the same solution holds along every subsequence of $\delta$ converging to $1$ (from below) so that the uniqueness of the equilibrium also follows. Finally, we can verify that there can be no other equilibria.

\bibliographystyle{ormsv080}
\bibliography{bargainbib}

\end{document}